\documentclass[aps,prd,amsmath,amssymb,twocolumn,superscriptaddress,preprintnumbers,nofootinbib]{revtex4-1}
\pdfoutput=1
\usepackage{graphicx}
\usepackage{amsthm}
\usepackage{amsmath}
\usepackage{graphicx,subfigure}% Include figure files
\usepackage{dcolumn}% Align table columns on decimal point
\usepackage{bm}% bold math
\usepackage{slashed}

\newcommand{\be}{\begin{eqnarray}}
\newcommand{\ee}{\end{eqnarray}}
\newcommand{\bea}{\begin{eqnarray}}
\newcommand{\eea}{\end{eqnarray}}

\newcommand{\GeV}{{~\rm GeV}}

\newcommand{\bi}{\begin{itemize}}
\newcommand{\ei}{\end{itemize}}
\newcommand{\benum}{\begin{enumerate}}
\newcommand{\eenum}{\end{enumerate}}

\begin{document}

\title{Baryogenesis through Neutrino Oscillations: A Unified Perspective}
\author{Brian Shuve}
\email{bshuve@perimeterinstitute.ca}
\affiliation{Perimeter Institute for Theoretical Physics 31 Caroline St. N, Waterloo, Ontario, Canada N2L 2Y5.}
\affiliation{Department of Physics \& Astronomy, McMaster University 1280 Main St. W. Hamilton, Ontario, Canada, L8S 4L8.}
\author{Itay Yavin}
\email{iyavin@perimeterinstitute.ca}
\affiliation{Department of Physics \& Astronomy, McMaster University 1280 Main St. W. Hamilton, Ontario, Canada, L8S 4L8.}
\affiliation{Perimeter Institute for Theoretical Physics 31 Caroline St. N, Waterloo, Ontario, Canada N2L 2Y5.}

%\date{\today}

\begin{abstract}
Baryogenesis through neutrino oscillations is an elegant mechanism which has found several realizations in the literature corresponding to different parts of the model parameter space. Its appeal stems from its minimality and dependence only on physics below the weak scale. In this work we show that, by focusing on the physical time scales of leptogenesis instead of the model parameters, a more comprehensive picture emerges. The different regimes previously identified can be understood as different relative orderings of these time scales. This approach also shows that all regimes require a coincidence of time scales and this in turn translates to a certain tuning of the parameters, whether in mass terms or Yukawa couplings. Indeed, we show that the amount of tuning involved in the minimal model is never less than one part in $10^5$. Finally, we explore an extended model where the tuning can be removed in exchange for the introduction of a new degree of freedom in the form of a leptophilic Higgs with a vacuum expectation value of the order of GeV.

\end{abstract}

\maketitle

%%%%%%%%%%%%%%%%
% Introduction
%%%%%%%%%%%%%%%%
\section{Introduction}
\label{sec:intro}

It is a theoretically attractive possibility to explain the baryon asymmetry of the Universe through the mechanism of sterile neutrino oscillations~\cite{Akhmedov:1998qx,Asaka:2005pn}.  The model is simple, containing only right-handed sterile neutrinos in addition to the Standard Model (SM). These neutrinos are light ($\sim$ GeV) and do not require any new physics at inaccessibly high energy scales. The model even holds the possibility that one of the sterile neutrinos can be the non-baryonic dark matter of the Universe~\cite{Asaka:2005pn,Asaka:2005an,Laine:2008pg,Canetti:2012kh}. Several past works have found different choices of the parameters that lead to the correct baryon asymmetry and identified several regimes~\cite{Asaka:2005pn,Shaposhnikov:2008pf,Asaka:2010kk,Canetti:2010aw,Asaka:2011wq,Drewes:2012ma,Canetti:2012kh}. 

It is the purpose of this work to present a unified perspective on leptogenesis through neutrino oscillations, weaving the  disjoint regimes previously identified in the literature into a single continuous picture. Our analysis focuses on the three important time scales for baryogenesis: the time of sterile neutrino oscillations, active-sterile neutrino equilibration time, and sphaleron decoupling time. We identify the different regimes as different relative orderings of these time scales and demonstrate the continuity of separate parts of the parameter space. This allows us to point out the most important effects contributing to the asymmetry in each regime. Along the way, we provide some improvement upon the calculation of the baryon asymmetry from neutrino oscillations by including the effects of scatterings between left-handed (LH) leptons and the thermal bath during asymmetry generation.

Aside from providing a unified framework, centering the discussion around the relevant time scales for baryogenesis brings to the forefront the need for an accidental coincidence between the different, unrelated scales in the problem. It is therefore no surprise that the framework is marred by the need for fine-tuning between its fundamental parameters in order to successfully generate the correct baryon asymmetry in the different regimes. We show that, while the fine-tuning present in different regimes appears in different parameters, the total fine-tuning is always at least at the level of $1/10^{5}$.

While possibly aesthetically unappealing, the persistent fine-tuning needed throughout the parameter space is not grounds to discount the framework. It is, however, possibly an invitation to explore extensions of the minimal model which would either alleviate some of the necessary tuning, or explain it as a small departure from a more symmetric phase. In the final part of this work, we therefore consider the possibility of an additional Higgs doublet with a small electroweak vacuum expectation value (VEV), which is coupled to all the leptons through larger-than-usual Yukawa couplings. We show that, aside from ameliorating the fine-tuning needed for successful leptogenesis, such a leptophilic Higgs doublet can be searched for directly and indirectly in high-energy reactions accessible at the LHC.

The paper is organized as follows: in Section \ref{sec:mechanism}, we review the qualitative features of baryogenesis through neutrino oscillations, providing a unified perspective on the physics responsible for leptogenesis in different regimes. In Section \ref{sec:models}, we discuss the different model parameters and their specific roles in leptogenesis. Section \ref{sec:calculation} presents the formalism for computing the baryon asymmetry and describes the modifications we make to the asymmetry evolution equations to account for SM thermal scatterings. In Section \ref{sec:BAU_minimal}, we confirm our unified perspective on leptogenesis with numerical studies of the different parameter regimes. We explicitly show the tuning required to obtain the observed baryon asymmetry throughout the different regimes of the minimal model. We then discuss the baryon asymmetry in a leptophilic Two Higgs Doublet Model (2HDM) in Section~\ref{sec:BAU_extended}, demonstrating how the asymmetry can be enhanced relative to the minimal model, eliminating any need for tuning. We also discuss experimental implications of this extended model. We close with a discussion and some remarks about the inclusion of dark matter in the model.

%%%%%%%%%%%%%%%%
% Mechanism
%%%%%%%%%%%%%%%%
\section{Review of Baryogenesis through Neutrino Oscillations}
\label{sec:mechanism}

It has been appreciated for some time that, in extensions of the Standard Model with sterile neutrinos, oscillations of the sterile neutrinos can be responsible for baryogenesis \cite{Akhmedov:1998qx,Asaka:2005pn}. We discuss the basic framework in this section, emphasizing the essential features that emerge from the model. We begin by presenting the model, and follow with a qualitative account of the asymmetry generation and its dependence on the underlying model parameter. 

\subsection{Neutrino Minimal Standard Model ($\nu$MSM)}

We  consider the Type I see-saw model for generating LH neutrino masses, also known as the {\bf Neutrino Minimal Standard Model} ($\nu$MSM). In this scenario, the SM is supplemented by three sterile neutrinos, $N_I$, with Majorana masses, $M_I$. In the basis where the charged lepton and sterile neutrino masses are diagonal, the Lagrangian is (see ref.~\cite{Drewes:2013gca} for a recent review)
\be\label{eq:typei}
\mathcal L_{\nu\mathrm{MSM}} = F_{\alpha I}\,L_\alpha \Phi N_I + (M_{N})_I\, N_I N_I,
\ee
where $\Phi$ is the electroweak doublet scalar responsible for giving mass to the SM neutrinos, and $L_\alpha$ is the SM lepton doublet of flavour $\alpha$.  When the scalar doublet develops a VEV, $\langle\Phi\rangle\ne0$, these interactions generate a small mass for the LH neutrinos through the see-saw mechanism~\cite{Minkowski:1977sc} of the order,
\be
\label{eq:seesaw_lag}
m_\nu &\sim& \frac{F^2 \,\langle\Phi\rangle^2}{M_N} \\ \nonumber
&\sim& 0.1\,\,\mathrm{eV} \left(\frac{\langle\Phi\rangle}{100\,\,\mathrm{GeV}}\right)^2\left(\frac{F}{10^{-7}}\right)^2 \left(\frac{\mathrm{GeV}}{M_N}\right).
\ee
Here, we used the parameters and masses most relevant for the current work:  a scalar VEV around the electroweak scale, $\langle\Phi\rangle\sim 100\GeV$; sterile-neutrino masses around or below the electroweak scale, $M_N \sim \GeV$; and small Yukawa couplings, $F\sim10^{-7}-10^{-8}$. With weak-scale sterile neutrino masses, the Yukawa couplings of the neutrino are only somewhat smaller than the electron Yukawa. 

With such a small coupling between LH leptons and sterile neutrinos, active-sterile neutrino scattering is out of equilibrium in the early universe and does not become rapid until $T\lesssim T_{\rm W}$, where $T_{\rm W}\approx 140$ GeV is the temperature of the sphaleron decoupling at the electroweak phase transition~\cite{Canetti:2012kh}. If there is a negligible concentration of sterile neutrinos immediately following inflation, then the sterile-neutrino abundance remains below its equilibrium value until it re-thermalizes at $T\lesssim T_{\rm W}$. As is well known, sphaleron processes active for $T>T_{\rm W}$ violate baryon number ($B$), but preserve the difference between baryon and lepton numbers ($B-L$), processing the  primordial lepton asymmetry into a baryon asymmetry \cite{Kuzmin:1985mm}. Thus, the final baryon asymmetry of the universe observed today is determined by the lepton asymmetry at the time of sphaleron decoupling, $T_{\rm W}$~\cite{D'Onofrio:2012jk}. In everything that follows, we therefore assume the physics associated with sphalerons to be present and concentrate on a detailed understanding of leptogenesis alone. 

\subsection{Sakharov Conditions for Leptogenesis}

The Sakharov conditions \cite{Sakharov:1967dj} necessary for generating a total lepton asymmetry are satisfied in the $\nu$MSM:
\benum
\item {\bf Violation of Standard Model lepton number:} The Yukawa coupling in Eq.~(\ref{eq:seesaw_lag}) preserves a generalized lepton number $L-N$ under which both SM and sterile neutrinos are charged. The $L-N$ symmetry is broken by the sterile neutrino Majorana mass, but rates of $(L-N)$-violating processes are suppressed by a factor of $M_N^2/T^2$ relative to $(L-N)$-preserving rates, and so total lepton number violation is generally ineffective for $T\gtrsim T_{\rm W}$. However, scattering processes $L_\alpha\rightarrow N_I^\dagger$ through the Yukawa interactions in Eq.~(\ref{eq:seesaw_lag}) violate \emph{Standard Model} lepton number, allowing the creation of equal asymmetries in $L$ and $N$ such that $L-N$ is still conserved. Sphalerons then convert the SM $L$ asymmetry into a baryon asymmetry.

\item {\bf $CP$ violation:} There are three $CP$ phases in the Yukawa couplings $F_{\alpha I}$. Together, these provide a sufficient source for leptogenesis through neutrino oscillations.

 \item {\bf Departure from equilibrium:} As discussed above, if $M_N\lesssim T_{\rm W}$, sterile neutrino scatterings are out of equilibrium provided there is no abundance of sterile neutrinos at the earliest times following inflation. Unlike many models of baryogenesis, the out-of-equilibrium condition is satisfied for an extended period in the early universe, with equilibration only occurring after sphaleron decoupling, $T\lesssim T_{\rm W}$.
\eenum

In the following subsection, we elaborate on the physical processes responsible for the production (and destruction) of the lepton asymmetry, and clarify which parameters  most strongly control the size of the baryon asymmetry. 

\subsection{Asymmetry Creation and Washout}
\label{sec:asymcreation}

%%%%%%%%%%%%%%%% FIGURE %%%%%%%%%%%%%%%%%%%%%
%\begin{figure}[t]
%\centering
%\includegraphics[width=0.4\textwidth]{plots/N_abundance.pdf}%
%\caption{Illustration of the evolution of the sterile neutrino abundance ($Y_N$) as a function of temperature. The plot shows a model with $M_N=1$ GeV, $F\approx10^{-6}$.}
%\label{fig:Nabundance}
%\end{figure}
%%%%%%%%%%%%%%%% FIGURE %%%%%%%%%%%%%%%%%%%%%
%
%Throughout, we quantify the abundances of a sterile neutrino labelled by $I$ as  the number density of that species, $n_I$ normalized to the entropy density, $Y_I = \frac{n_I}{s}$.

\begin{figure}[t]
\centering
\includegraphics[scale=0.7]{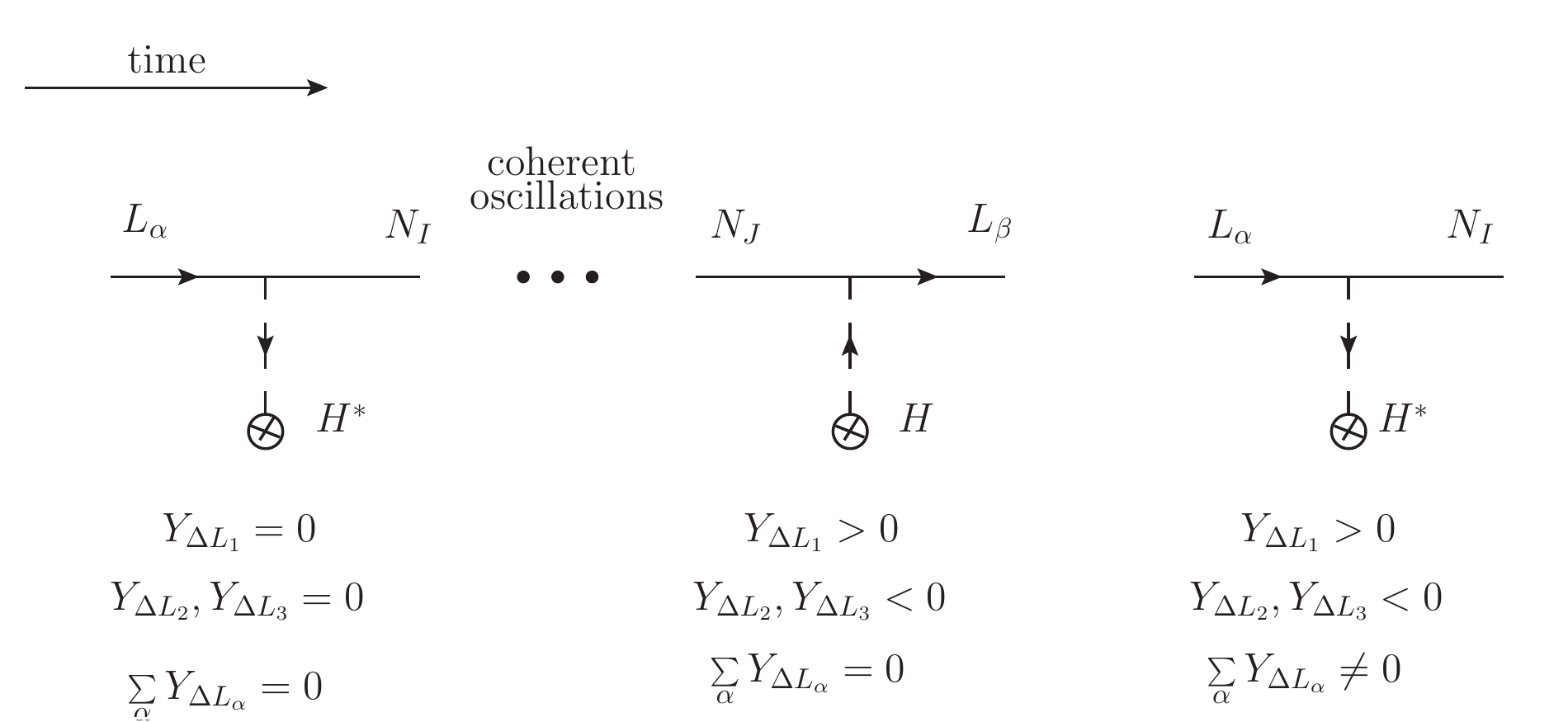}%
\caption{The basic stages leading to the creation of a total lepton asymmetry from left to right:  out-of-equilibrium scattering of LH leptons begin to populate the sterile neutrino abundance at order $\mathcal{O}(|F|^2)$;  after some time of coherent oscillation, a small fraction of the sterile neutrinos scatter back into LH leptons to create an asymmetry in individual lepton flavours at order $\mathcal{O}(|F|^4)$; finally, at order $\mathcal{O}(|F|^6)$, a total lepton asymmetry is generated due to a difference in scattering rate into sterile neutrinos among the different active flavours.}
\label{fig:schematic}
\end{figure}

The basic stages leading to the creation of a total SM lepton asymmetry are shown in Fig.~\ref{fig:schematic}. Immediately following inflation, there is no abundance of sterile neutrinos, and out-of-equilibrium scatterings mediated by the Yukawa couplings begin to populate the sterile sector, as shown on the left side of Fig.~\ref{fig:schematic}. The sterile neutrinos are produced in a coherent superposition of mass eigenstates\footnote{This is true assuming generic parameters with no special alignment of the sterile-neutrino interaction and mass eigenstates.} and remain coherent as long as the active-sterile Yukawa coupling remains out of equilibrium, since in the minimal model there are no other interactions involving the sterile neutrinos.

%\addvspace{2\baselineskip}
\onecolumngrid

%%%%%%%%%%%%%%% FIGURE %%%%%%%%%%%%%%%%%%%%%
%%%%%%%%%%%%%%% FIGURE %%%%%%%%%%%%%%%%%%%%%

\hrulefill
\twocolumngrid

Some time later, a subset of the sterile neutrinos scatter back into LH leptons, mediating $L_\alpha\rightarrow L_\beta$ transitions as shown in the centre of Fig.~\ref{fig:schematic}. Since the sterile neutrinos remain in a coherent superposition in the intermediate time between scatterings, the transition rate $L_\alpha\rightarrow L_\beta$ includes an interference between propagation mediated by the different sterile neutrino mass eigenstates. The different mass eigenstates have different phases resulting from time evolution; for sterile neutrinos $N_I$ and $N_J$, the relative phase accumulated during a small time $dt$ is $e^{-i(\omega_I-\omega_J)\,dt}$, where 
\be
\omega_I-\omega_J \approx \frac{(M_N)_I^2-(M_N)_J^2}{2T} \equiv \frac{(M_N)_{IJ}^2}{2T}.
\ee
In the interaction basis, this $CP$-even phase results from an oscillation between different sterile neutrino flavours, and explains the moniker of leptogenesis through neutrino oscillations. 

When combined with the $CP$-odd phases from the Yukawa matrix, neutrino oscillations lead to a difference between the $L_\alpha\rightarrow L_\beta$ rate and its complex conjugate,
\small
\be\label{eq:flavour_rate_difference}
\Gamma(L_\alpha\rightarrow L_\beta)-\Gamma(L_\alpha^\dagger\rightarrow L_\beta^\dagger) &\propto& \sum_{I\neq J}\mathrm{Im}\left[\exp\left(-i\int_0^t\frac{M_{IJ}^2}{2T(t')}dt'\right)\right] \nonumber  \\ 
&\times&\mathrm{Im}\left[F_{\alpha I}F_{\beta I}^*F^*_{\alpha J}F_{\beta J}\right].
\ee
\normalsize
In the absence of efficient washout interactions, which is ensured by the out-of-equilibrium condition, this difference in rates creates asymmetries in the individual LH lepton flavours $L_\alpha$.

Denoting the individual LH flavour abundances (normalized by the entropy density, $s$) by $Y_{L_\alpha} \equiv n_{L_\alpha}/s$ and the asymmetries by $Y_{\Delta L_\alpha} \equiv Y_{L_\alpha} - Y_{L^\dagger_\alpha}$, we note that the processes at order $\mathcal{O}(|F|^4)$ discussed thus far only convert $L_\alpha$ into $L_\beta$, conserving \emph{total} SM lepton number,
\be
\label{eq:no_asym_order_F4}
Y_{\Delta L_{\rm tot}} = \sum_\alpha\, Y_{\Delta L_\alpha} = 0 \quad \quad \quad \text{at}~~ \mathcal{O}(|F|^4).
\ee
Since sphalerons couple to the total SM lepton number, it follows that no baryon asymmetry is generated at this order as well, $Y_{\Delta B_{\rm tot}} =0$. Total lepton asymmetry is, however, generated at order $\mathcal{O}(|F|^6)$: the excess in each individual LH lepton flavour due to the asymmetry from Eq.~(\ref{eq:flavour_rate_difference}) leads to a slight increase of the rate of $L_\alpha\rightarrow N^\dagger$ vs.~$L_\alpha^\dagger\rightarrow N$. The result is that active-sterile lepton scatterings can convert individual lepton flavour asymmetries into asymmetries in the sterile neutrinos. But, since the rates of conversion, $\Gamma(L_\alpha\rightarrow N^\dagger)$, are generically different for each lepton flavour $\alpha$, this leads to a depletion of some of the individual lepton asymmetries at a faster rate than others, leading to an overall SM lepton asymmetry and an overall sterile neutrino asymmetry. Because $L_{\rm tot}-N$ is conserved for $T\gg m_N$, this gives
\be\label{eq:totalbaryon}
\frac{dY_{\Delta N_{\rm tot}}}{dt} = \frac{dY_{\Delta L_{\rm tot}}}{dt} = \sum_{\alpha,I}\, Y_{\Delta L_\alpha}\,\Gamma(L_\alpha \rightarrow N_I^\dagger).
\ee
Therefore, while the sum of the LH lepton flavour asymmetries vanishes at $\mathcal{O}(|F|^4)$ as in Eq.~(\ref{eq:no_asym_order_F4}), the fact that the LH leptons scatter at different rates into the sterile sector results in a non-vanishing total lepton asymmetry at $\mathcal{O}(|F|^6)$.

The asymmetry generated at early times can be destroyed later  when the different lepton flavours establish chemical equilibrium with the sterile neutrinos. This occurs when the transition rate exceeds the Hubble rate, $\Gamma(L_\alpha\rightarrow N^\dagger)\sim (FF^\dagger)_{\alpha\alpha}\,T \gtrsim H$. At this time, the particular lepton flavour $L_\alpha$ reach chemical equilibrium with the sterile neutrinos, but not yet with the other SM leptons since lepton flavour is still conserved to a good approximation. Suppose, for concreteness, that $L_\tau$ comes into equilibrium with the sterile neutrinos, but $L_\mu$ and $L_e$ remain out of equilibrium. Then, {\bf before $L_\tau$ comes into equilibrium}, we can use the  approximate conservation of $\Delta L_{\rm tot}-\Delta N=0$  to write
\bea
\nonumber
Y_{\Delta L_e} &\equiv& x, \quad Y_{\Delta L_\mu} \equiv y,  \quad Y_{\Delta L_\tau} \equiv z,
\\ \nonumber
\\ \label{eq:before_eq}
Y_{\Delta N} &=& Y_{\Delta L_{\rm tot}} = x+y+z.
\eea
Once $L_\tau$ comes into equilibrium with the sterile neutrinos, the Yukawa coupling $F_{\tau I}\,L_\tau H N_I$ leads to the equilibrium chemical potential relation\footnote{Strictly speaking, the relation is $\mu_{L_\tau}+\mu_\Phi = -\mu_N$; for illustrative purposes, and because $\mu_\Phi$ is typically small compared to the LH SM lepton flavour asymmetries, we set $\mu_\Phi=0$ here.} $\mu_{L_\tau} = -\mu_{N}$. Because the fields $L_\tau$ and $N$ both have two components and opposite-sign chemical potentials, their asymmetries are therefore equal and opposite: $Y_{\Delta L_\tau}=-Y_{\Delta N}$. The flavour asymmetries $\Delta L_\mu$ and $\Delta L_e$  remain unchanged because they are not yet in equilibrium. Together with the fact that the $\Delta L_{\rm tot}-\Delta N\approx0$ at high temperature, this implies that {\bf after $L_\tau$ equlibration:}
\be
Y_{\Delta N} = Y_{\Delta L_{\rm tot}} = \frac{x+y}{2}.\label{eq:after_eq}
\ee
Therefore, at the point of equilibration of the $\tau$ flavour, the total lepton asymmetry rapidly changes from $x+y+z$ to $(x+y)/2$, and the $\Delta L_\tau$ asymmetry rapidly changes from $z$ to $-(x+y)/2$. We see that nothing remains of the original $\Delta L_\tau$ flavour asymmetry $z$ by the time the tau flavour comes fully into equilibrium. Both the total lepton asymmetry and $L_\tau$ flavour asymmetry  change dramatically when equilibrium is reached, with the asymmetry after equilibrium being fixed by the remaining  asymmetries in the flavours $L_e$ and $L_\mu$. It is also apparent that if all of the flavours come into equilibrium,  $\Delta L_e=\Delta L_\mu=\Delta L_\tau=\Delta N = 0$. Clearly,  sphaleron interactions must decouple before this time is reached or no baryon asymmetry results from sterile neutrino oscillation.

\subsection{Qualitative Dependence of Asymmetry on Model Parameters}

In the discussion above we have highlighted three important elements for generating a non-zero baryon asymmetry: the coherent oscillation and interference of different sterile neutrino states in $L_\alpha\rightarrow L_\beta$ scattering; the magnitudes of the Yukawa couplings which determine the rates of processes generating the lepton asymmetry; and the presence of differences in scattering rates of individual LH lepton flavours into sterile neutrinos. We now elaborate on each, as they have implications for what parts of parameter space maximize the baryon asymmetry in the minimal model, and how new interactions can enhance the asymmetry. \\

%%%%%%%%%%%%%%%% FIGURE %%%%%%%%%%%%%%%%%%%%%
%\begin{figure}[t]
%\centering
%\includegraphics[width=0.35\textwidth]{plots/asym_deltaM.pdf}\hspace{1.5cm}\includegraphics[width=0.35\textwidth]{plots/asym_yukawa.pdf}%
%\caption{{\bf Do we want to include these plots? They show how the asymmetry evolves with time for different $\Delta M$, Yukawa couplings. If so, re-do 
%with nicer colours and consistent notation with other plots.} }
%\label{fig:asymmetry_parametric_dependence}
%\end{figure}
%%%%%%%%%%%%%%%% FIGURE %%%%%%%%%%%%%%%%%%%%%

\noindent {\bf Sterile neutrino mass splitting:} In the absence of a sterile-neutrino mass splitting, the sterile-neutrino masses and couplings can be simultaneously diagonalized, and there is no coherent oscillation/interference as required in Fig.~\ref{fig:schematic}. The size of the mass splitting dictates the time scale at which the phases of the coherently evolving sterile neutrino eigenstates become substantially different. From Eq.~(\ref{eq:flavour_rate_difference}) and the Hubble scale in a radiation-dominated universe \cite{Kolb:1990vq},
\be
H = \frac{1.66\sqrt{g_*}\,T^2}{M_{\rm Pl}}
\ee
($g_*$ is the number of relativistic degrees of freedom), we have an $\mathcal{O}(1)$ phase from oscillation at
\be
t_{\rm osc} \approx \left(\frac{3\sqrt{M_{\rm Pl}}}{1.66 g_*^{1/4} \sqrt{2}\left(M_{N3}^2-M_{N2}^2\right)}\right)^{2/3}.
\ee
The oscillation time is later for smaller mass splittings. At later times, the rates of scattering between active-sterile neutrinos is larger relative to the Hubble scale, and the asymmetry is consequently larger. Therefore,  small but non-zero sterile neutrino mass splittings enhance the size of the baryon asymmetry \cite{Akhmedov:1998qx,Asaka:2005pn}, as long as  active-sterile neutrino scattering is not so rapid as to decohere the sterile neutrinos prior to coherent oscillation. At even later times, $t\gg t_{\rm osc}$, the oscillation rate is rapid compared to the Hubble scale, and sterile neutrinos produced at different times have different phases; averaging over the entire ensemble results in a cancellation of the asymmetry production from each sterile neutrino. Therefore, the lepton flavour asymmetries are dominated by production at $t\sim t_{\rm osc}$.\\

\noindent {\bf Magnitude of Yukawa couplings:} The rate of production of individual lepton flavour asymmetries in (\ref{eq:flavour_rate_difference}) is $\mathcal{O}(|F|^4)$; therefore, increasing the magnitude of the Yukawa couplings gives a substantial enhancement to the individual flavour asymmetries. Also, larger Yukawa couplings give a more rapid transfer rate from the individual lepton flavour asymmetries into a total lepton asymmetry at order $\mathcal{O}(|F|^6)$, further enhancing the baryon asymmetry. However, increase in the Yukawas also enhances the washout processes. The characteristic time scale associated with the washout of lepton flavour $\alpha$ is
\be
t_{\alpha\,\mathrm{washout}} \sim \frac{1}{\Gamma(L_\alpha \rightarrow N^\dagger)} \sim \frac{1}{(FF^\dagger)_{\alpha\alpha}T}.
\ee
 If the Yukawa coupling is too large, then washout occurs before the electroweak phase transition,  and all lepton flavour asymmetries are driven to zero in the equilibrium limit. The final asymmetry is maximal when the Yukawa couplings are large enough to equilibrate all lepton species immediately after the electroweak phase transition, but not larger. Since sphalerons decouple at $T_{\rm W}$, the baryon asymmetry is frozen in even though the lepton asymmetry is rapidly damped away shortly after the phase transition. This is true, provided that the Yukawa couplings are not so large that the sterile neutrinos decohere before the time of asymmetry generation, which can occur if the oscillation time is very late ($\Delta M_N/M_N\ll1$).\\

\noindent {\bf Lepton flavour dependence in scattering rates:} As discussed in Section \ref{sec:asymcreation}, the generation of individual flavour asymmetries at order $\mathcal{O}(|F|^4)$ due to sterile neutrino oscillations is insufficient. A total SM lepton asymmetry is generated only at $\mathcal{O}(|F|^6)$ due to flavour-dependent scattering rates according to Eq.~(\ref{eq:totalbaryon}). In the absence of lepton flavour-dependent effects, the individual scattering rates are all equal, $\Gamma(L_e\rightarrow N^\dagger) = \Gamma(L_\mu \rightarrow N^\dagger) = \Gamma(L_\tau \rightarrow N^\dagger)$, and the total lepton asymmetry remains zero even at higher orders in $F$. Therefore, differences in lepton flavour rates are crucial to generate a baryon asymmetry. Fortunately, there is already evidence for lepton flavour dependence in interactions with neutrinos. First, the structure of the LH neutrino masses and mixing angles tells us that their Yukawa couplings are non-universal, and proportional to the mixing angles $\theta_{ij}$. Second, the $CP$ phases appear in very particular terms in the interaction: for instance, the Dirac phase $\delta$ appears only in terms proportional to $\sin\theta_{13}$. Changing the phase can lead to constructive or destructive interference of rates involving a specific lepton flavour. The total lepton asymmetry is maximized in regions of parameter space that accentuate the differences between lepton flavour interaction rates. The importance of flavour effects was recently emphasized in \cite{Drewes:2012ma}.

%%%%%%%%%%%%%%% FIGURE %%%%%%%%%%%%%%%%%%%%%
\begin{figure}[t]
\centering
\includegraphics[width=0.39\textwidth]{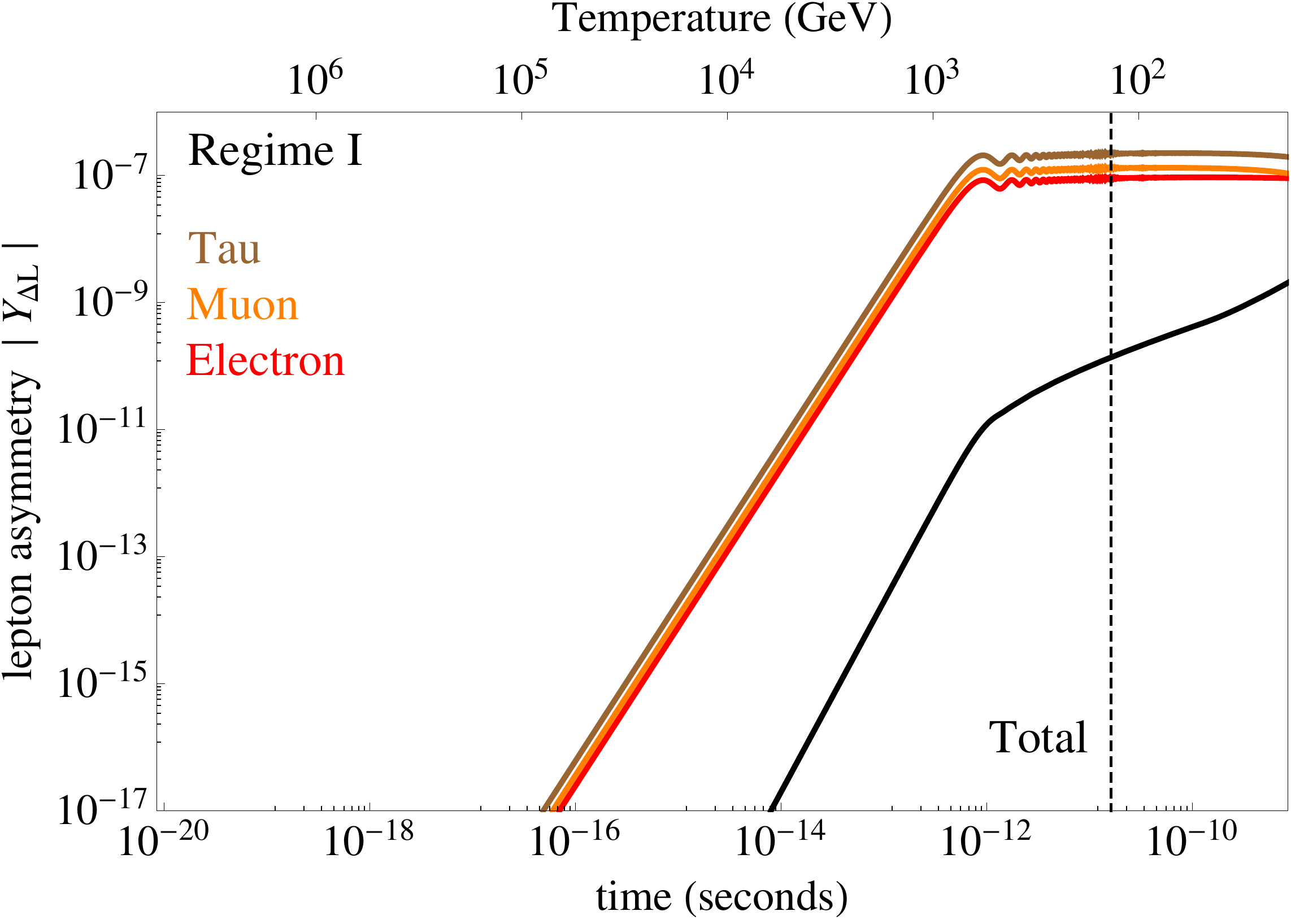}
\includegraphics[width=0.39\textwidth]{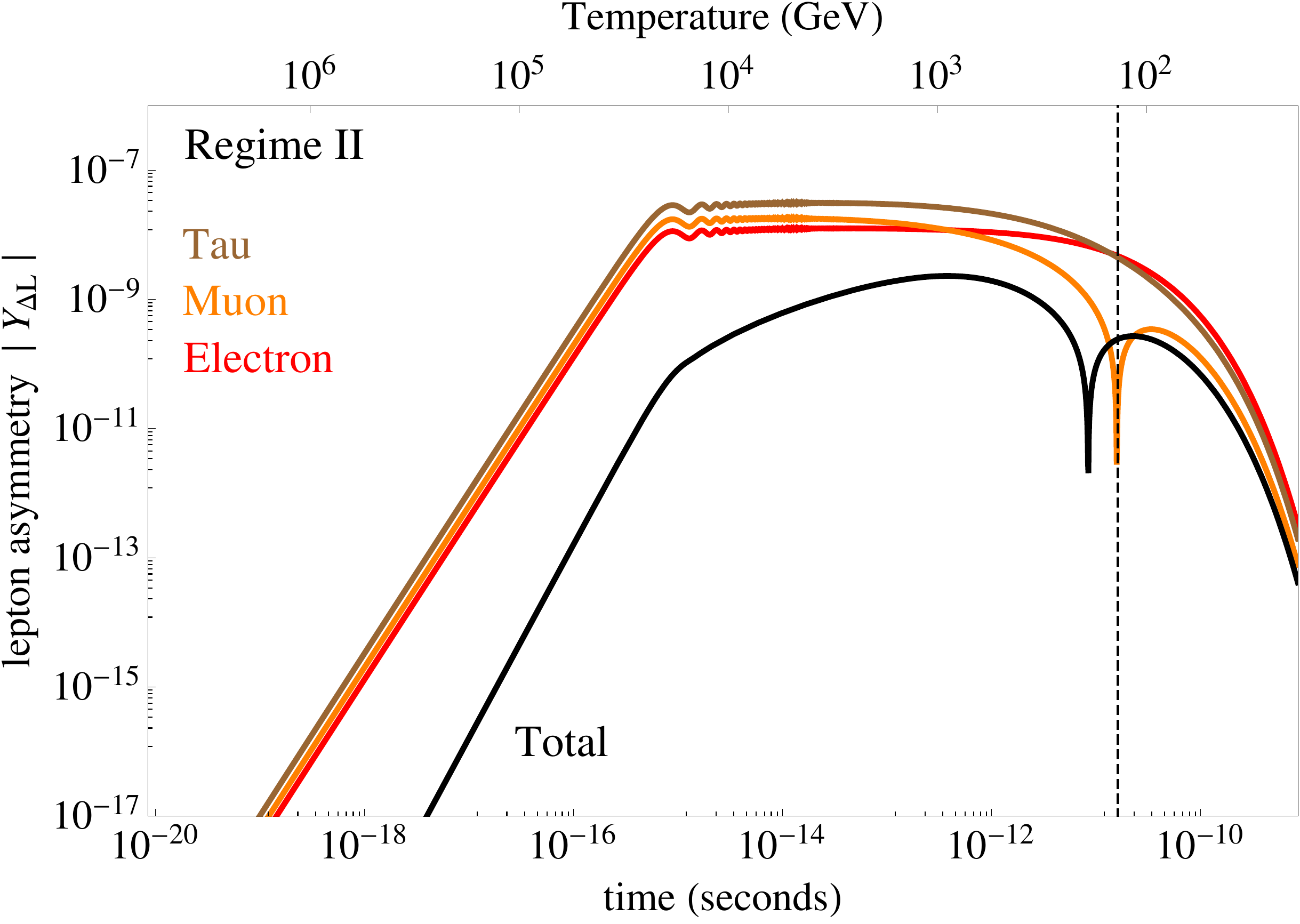}
\includegraphics[width=0.39\textwidth]{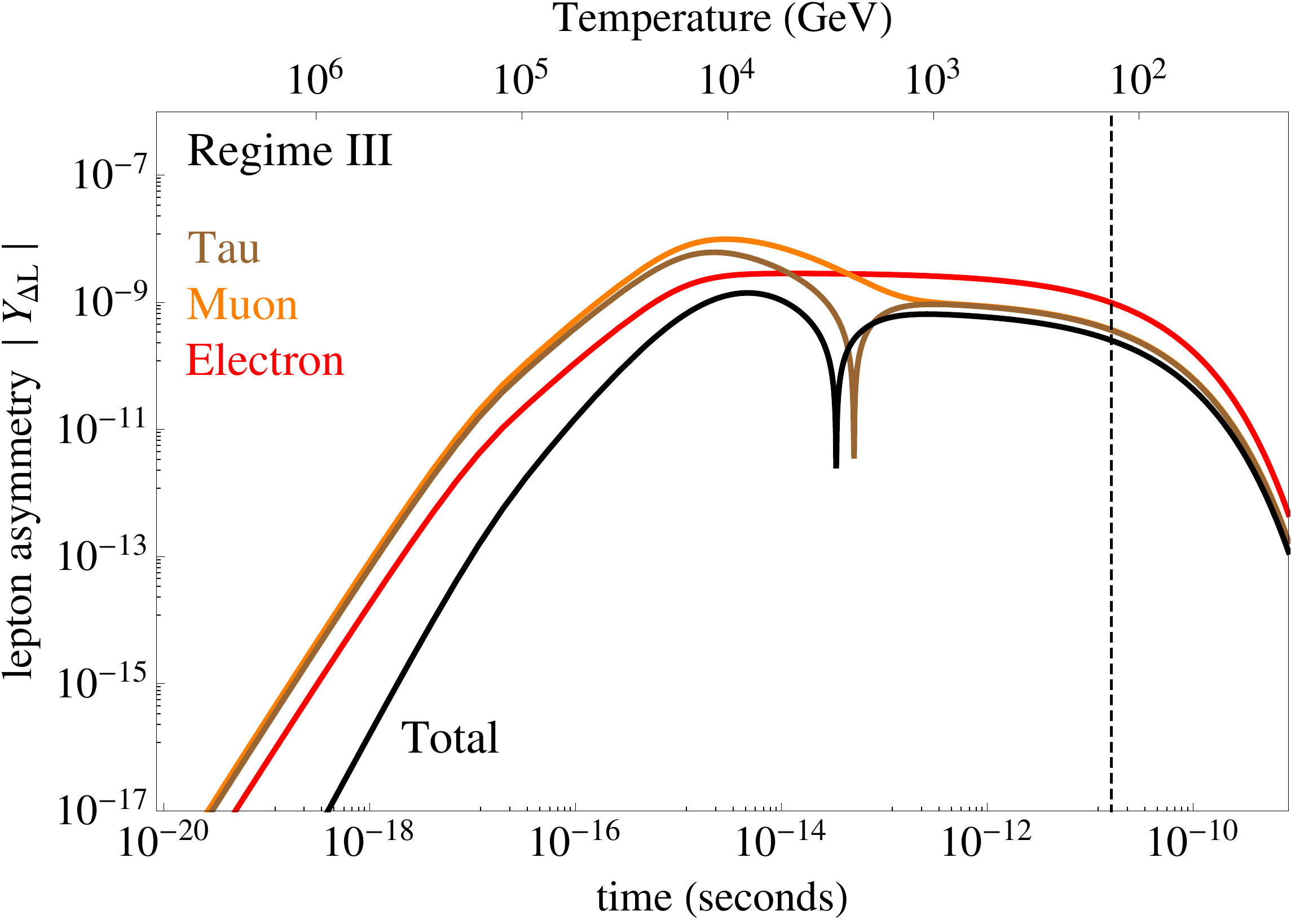}
\caption{Time evolution of the individual lepton flavour asymmetries and the total lepton asymmetry in different parameter regimes. The dashed vertical line indicates the electroweak phase transition. {\bf (Top)} Regime I: The Yukawa couplings are small enough that washout effects are always irrelevant; {\bf (Centre)} Regime II: The Yukawa couplings are large enough that equilibration of the sterile neutrinos occurs at the electroweak phase transition; {\bf (Bottom)} Regime III: Large lepton flavour dependence in the $L\rightarrow N^\dagger$ rates, such that $\Gamma(L_e\rightarrow N^\dagger)\ll \Gamma(L_\mu\rightarrow N^\dagger)$, $\Gamma(L_\tau\rightarrow N^\dagger)$. The Yukawa couplings are even larger than in Regime II, such that $L_\tau$ and $L_\mu$ equilibrate completely with the sterile neutrinos at $T\gg T_{\rm W}$, and $L_e$ comes into equilibrium at $T\sim T_{\rm W}$.}
\label{fig:asymmetry_regimes}
\end{figure}
%%%%%%%%%%%%%%% FIGURE %%%%%%%%%%%%%%%%%%%%%

Considering all of these effects, we identify regimes of parameter space depending on the relative time scales of oscillation, sterile neutrino equilibration, and sphaleron decoupling. Because these are the only time scales in the problem, this results in three regimes that completely characterize the minimal model. {\bf Regime I} is defined by $t_{\rm osc}\lesssim t_{\rm W}\ll t_{\rm eq}$ and was considered in the original works on baryogenesis from neutrino oscillations \cite{Akhmedov:1998qx,Asaka:2005pn}. Equilibration occurs long after the electroweak phase transition and so washout effects are entirely irrelevant. However, the total baryon asymmetry is also generally too small because the small Yukawa couplings implied by the late washout also suppress the rates in Eq.~(\ref{eq:totalbaryon}). To account for the observed baryon asymmetry, the oscillation time scale must be made as large as possible and thus close to the electroweak scale, $t_{\rm osc}\lesssim t_{\rm W}$. This coincidence of scales is achieved through a fine-tuning of the sterile-neutrino mass splitting $\Delta M_N\equiv M_{N_3}-M_{N_2}\lesssim(10^{-6}-10^{-8})M_N$. In {\bf Regime II}, the sterile neutrinos come into equilibrium around the weak scale, and the sphaleron processes freeze out ``just in time'' to avoid washing out the entire asymmetry $t_{\rm osc}\ll t_{\rm eq}\sim t_{\rm W}$. This alleviates some of the tuning necessary in the mass splitting $\Delta M_N$, but the coincidence between the equilibration scale and the electroweak scale requires some tuning in the Yukawa couplings. Finally, in {\bf Regime III}, first identified in \cite{Drewes:2012ma}, the Yukawa couplings are made even larger. This generically results in the equilibration time, $t_{\rm eq}$, being even earlier than the weak scale. This would have been a phenomenological disaster with the entire asymmetry being washed-out too early, except that destructive interference in the scattering rate for one of the lepton flavours makes it much smaller than the others. The result is a time scale ordering $t_{\rm osc}\ll t_{\rm{eq},\,\alpha} \ll t_{\rm{eq},\,\beta}\approx t_{\rm W} $. In this regime, even less tuning is necessary in the mass splitting, but the unnaturally early equilibration time for two of the flavours requires a substantial tuning in the Yukawa couplings. We quantify this combined tuning below in section~\ref{sec:tuning}.

In Fig.~\ref{fig:asymmetry_regimes}, we show the time evolution of the individual lepton flavour asymmetries and total lepton asymmetry for each of the regimes discussed above. In Regime II, the downward spikes in the total lepton asymmetry and the $L_\mu$ asymmetry are the result of a change in sign of the asymmetry, as shown in Eqs.~(\ref{eq:before_eq})-(\ref{eq:after_eq}) and the related discussion. The spike indicates the time of the equilibration of $L_\mu$ around $T=200\,\,\mathrm{GeV}\sim T_{\rm W}$. In Regime III, these spikes occur due to $L_\mu$ and $L_\tau$ equilibration around $T\gtrsim3$ TeV. Indeed, the hallmark of Regime III is that the equilibration temperatures of $L_\mu$ and $L_\tau$ are more than an order of magnitude larger than that of $L_e$ at $T\lesssim100$ GeV, whereas in Regime II, all flavours come into equilibrium near the same scale.

%%%%%%%%%%%%%%%%
% Models
%%%%%%%%%%%%%%%%
\section{Model Parameterization}
\label{sec:models}

 The model we consider is the $\nu$MSM \cite{Asaka:2005pn}, with the Lagrangian given in Eq.~(\ref{eq:typei}). At low temperatures $T\ll M_I$, the SM neutrinos acquire a mass in the effective theory,
\be\label{eq:seesaw}
\left(m_{\nu}\right)_{\alpha\beta} = \langle \Phi\rangle^2 \left(F M_N^{-1} F^{\rm T}\right)_{\alpha\beta}.
\ee
This is the usual see-saw suppression of the SM neutrino masses, and its parametric scaling was discussed in relation with Eq.~(\ref{eq:seesaw_lag}) . The observed masses and mixings of the SM neutrinos are \cite{Beringer:1900zz},
\bi
\item $|\Delta m_{\rm atm}^2|=2.35^{+0.12}_{-0.09}\times10^{-3}\,\,\mathrm{eV}^2$,\\ ~\\ $m_{\rm sol}^2 = 7.58^{+0.22}_{-0.26}\times10^{-5}\,\,\mathrm{eV}^2$, \\~\\ and $\sum_i m_{\nu_i}\lesssim$ eV.
\item $\sin^2\theta_{12}=0.312^{+0.018}_{-0.015}$, \\ ~\\$\sin^2\theta_{23}=0.42^{+0.08}_{-0.03}$, \\~\\and $\sin^22\theta_{13} = 0.096\pm0.013$.
\ei
The data are consistent with one of the LH neutrinos being massless, and one of the sterile neutrinos being largely decoupled from the SM. This decoupled sterile neutrino, which we take for concreteness to be $N_1$, is a possible dark matter candidate, but does not play a role in leptogenesis. Therefore, we consider an effective theory with only $N_2$ and $N_3$ as the two sterile neutrinos.

The assumption of one massless LH neutrino fixes the other LH neutrino masses, up to a discrete choice of mass hierarchy. We consider the normal hierarchy, where $m_{\nu 1}=0$, $m_{\nu 2}= \sqrt{m_{\rm sol}^2}\approx 9$ meV, and $m_{\nu 3} = \sqrt{|\Delta m_{\rm atm}^2|}\approx 49$ meV. It is also possible for the set-up to be realized in the inverted hierarchy ($m_3=0$, $m_2\sim m_1$), but the qualitative dependence of the baryon asymmetry on  model parameters is similar to the normal hierarchy, while the value of the baryon asymmetry can be somewhat larger with an inverted hierarchy \cite{Canetti:2010aw}. We focus exclusively on the normal hierarchy as a benchmark, since our results also qualitatively hold in the inverted hierarchy and easily generalize to that scenario.

The magnitudes of the Yukawa couplings $F_{\alpha2}$, $F_{\alpha3}$ are crucial for successful baryogenesis. To better understand the connection between the Yukawa couplings and physical parameters, we decompose the Yukawa couplings with the Casas-Ibarra parameterization~\cite{Casas:2001sr},
\be\label{eq:casasibarra}
F = \frac{i}{\langle \Phi\rangle} U_\nu \, \sqrt{m_\nu}\, R^*\, \sqrt{M_N}.
\ee
Here $U_\nu$ is the MNS matrix containing mixing angles and $CP$ phases from the LH neutrino mixing~\cite{Maki:1962mu}, $m_\nu$ is a diagonal matrix of LH neutrino masses, and $M_N$ is a diagonal matrix of sterile neutrino masses. The matrix $R$ is an orthogonal matrix specifying the mixing between sterile neutrino mass and interaction eigenstates; in the case of a normal neutrino hierarchy with two sterile neutrinos, $R$ has the form
\be
R_{\alpha I} = \left(\begin{array}{cc} 0 & 0 \\
\cos\omega & -\sin\omega \\
\sin\omega & \cos\omega \end{array}\right),
\ee
where $\omega$ is a complex angle parameterizing the misalignment between sterile neutrino mass and interaction eigenstates. We refer the reader to Appendix~\ref{sec:casasibarra} for the explicit form of the decomposition in Eq.~(\ref{eq:casasibarra}). We note that when $\omega=0$, the Yukawa interactions can be diagonalized in the sterile neutrino mass basis, and the interference/oscillation necessary for leptogenesis is absent. 

The parameters that emerge out of the decomposition of the Yukawa couplings can be grouped as follows:
\bi
\item \emph{Parameters currently fixed by experiment:}
\bi
\item Three LH neutrino masses, $(m_\nu)_i$;
\item Three LH neutrino mixing angles from the MNS matrix: $\theta_{12}$, $\theta_{13}$, $\theta_{23}$.
\ei
\item\emph{Parameters constrained by experiment:}
\bi
\item The VEV $\langle\Phi\rangle$. If $\Phi$ is the SM Higgs, it has the value $\langle\Phi\rangle=174$ GeV; otherwise, it is undetermined but constrained by bounds on new sources of electroweak symmetry breaking (see Section \ref{sec:experiments}).
\ei
\ei

\vspace{5mm}

\bi
\item \emph{Unconstrained parameters:}
\bi
\item Two phases from LH neutrino mixing: a Majorana phase, $\eta$, and a Dirac phase, $\delta$;
\item Two sterile neutrino masses, $(M_N)_I$;
\item A complex mixing angle, $\omega$.
\ei

\ei
These parameters, in turn, influence the dynamics of baryogenesis in three key ways: by setting the magnitude of the Yukawa couplings, controlling lepton flavour dependence in scattering rates, and providing the $CP$ violation necessary for baryogenesis. We consider each in turn.\\

\noindent{\bf Magnitude of Yukawa couplings:} The  scaling of the Yukawa magnitude is generally fixed by the see-saw relation $m_\nu \sim FF^{\rm T}\langle\Phi\rangle^2/m_N$. In particular, increasing either the sterile or LH neutrino masses enhances the Yukawa couplings, as does decreasing $\langle\Phi\rangle$. The presence of the complex parameter $\omega$, however, allows for an enhancement of the Yukawas well beyond the na\"ive see-saw value. The reason is that, while $FF^{\rm T}$ is fixed by the LH neutrino masses, physical rates depend on the quantities $FF^{\dagger}$ and $F^\dagger F$. It is therefore possible that squaring the complex terms in $\omega$ leads to large cancellations among the terms in $F$ such that $FF^{\rm T}\ll FF^\dagger$. The Yukawa couplings are actually enhanced \emph{exponentially} by the imaginary part of $\omega$ as
\be
|F|^2 \propto \frac{m_{\nu}M_N}{\langle\Phi\rangle^2} \cosh(2\mathrm{Im}\,\omega)
\ee
when $|\mathrm{Im}\,\omega|$ is large. The parameter $\mathrm{Im}\,\omega$ does not otherwise have any impact on experimentally observed quantities, and it can be thought of as a dial to enhance the rates of sterile neutrino production and scattering. 

When $|\mathrm{Im}\,\omega|\gg1$, the see-saw relation only holds due to a precise cancellation of parameters, and  there is a very specific alignment of the Yukawa couplings associated with this enhancement. While the Yukawa matrix is stable under radiative corrections, and therefore technically natural in the sense of 't Hooft \cite{'tHooft:1979bh}, the physically observed parameters (such as the LH neutrino masses) change significantly under small perturbations of the Yukawa matrix entries. This results in tuning in the sense of Barbieri and Giudice \cite{Barbieri:1987fn},  which is quantified by observing how the physical masses $m_\nu$ change under perturbations of the Yukawa coupling, $F$. For concreteness, consider a Yukawa matrix decomposed according to eq.~(\ref{eq:casasibarra}), which is then perturbed according to $F_{22}\rightarrow (1+\epsilon)F_{22}$ but otherwise left unchanged. In the simplest case with $\theta_{\alpha I} = \mathrm{Re}\,\omega=0$, the LH neutrino masses have a simple analytic form and the eigenvalue $m_2$ changes according to
\be\label{eq:masstuning}
\frac{d\log m_2}{d \epsilon} = 1+\cosh(2\mathrm{Im}\,\omega).
\ee
With non-zero $\theta$ and $\mathrm{Re}\,\omega$, the change in LH neutrino masses from $\epsilon$ is apportioned among the different mass eigenstates, but the overall shift is of the order of eq.~(\ref{eq:masstuning}). We have also verified this numerically. Therefore, while it is possible to exponentially enhance the rates relevant for baryogenesis in the minimal sterile neutrino model, it necessarily implies an exponential tuning of the Yukawa couplings to obtain the observed LH neutrino masses. Since much of the viable parameter space in the baryogenesis studies of \cite{Canetti:2010aw,Drewes:2012ma,Canetti:2012kh} requires $|\mathrm{Im}\,\omega|\gg1$, baryogenesis in these set-ups is unnatural in the sense of ref.~\cite{Barbieri:1987fn}.\\

\noindent{\bf Lepton flavour-dependent scattering rates:} As discussed in Section \ref{sec:mechanism}, lepton flavours must have different scattering rates into sterile neutrinos in order to convert the individual lepton flavour asymmetries into a total lepton asymmetry. There are several parameters in $F$ that contribute differently to the asymmetry generation and scattering rates for each flavour. The LH neutrino mass hierarchy ($m_1\ll m_2\ll m_3$) and hierarchy among mixing angles ($\theta_{13}\ll\theta_{12}\sim\theta_{23}$) provides some differentiation among lepton flavours; this typically suppresses rates of electron scattering versus the corresponding rates for muons and taus.

The $CP$ phases $\delta$ and $\eta$ also play an important role in distinguishing lepton flavours. Since they appear in different entries of the MNS matrix, these phases can lead to constructive or destructive interference among processes involving specific lepton flavours. For instance, when $\delta+\eta=-\pi/2$, there is  destructive interference between the $L_e\rightarrow N_2^\dagger$ and $L_e\rightarrow N_3^\dagger$ amplitudes, resulting in $\Gamma(L_e\rightarrow N^\dagger) \ll \Gamma(L_\mu\rightarrow N^\dagger)\sim\Gamma(L_\tau\rightarrow N^\dagger)$ \cite{Asaka:2011pb,Drewes:2012ma}. Such interference effects can substantially alter the rates of asymmetry creation or washout, modifying the total baryon asymmetry. We emphasize that \emph{this effect is independent of the role of the phases in $CP$ violation.} In fact, over many regions of  parameter space, this is the dominant contribution of the Majorana and Dirac $CP$ phases to the baryon asymmetry.\\

\noindent{\bf $CP$ violation:} There are three sources of $CP$ violation in the theory: the Majorana phase $\eta$, the Dirac phase $\delta$, and the complex sterile neutrino mixing angle $\omega$. Its imaginary part, $\mathrm{Im}\,\omega$, gives rise to two phases that are related to one another,
\bea
\label{eqn:tan_phi1}
\tan\phi_1 &\equiv& \tan(\arg\cos\omega)=-\tan(\mathrm{Re}\,\omega)\tanh(\mathrm{Im}\,\omega),\\
\label{eqn:tan_phi2}
\tan\phi_2 &\equiv& \tan(\arg\sin\omega)=\cot(\mathrm{Re}\,\omega)\tanh(\mathrm{Im}\,\omega).
\eea
According to the Sakharov conditions, a source of $CP$ violation is required to generate a baryon asymmetry. Except where $\delta$, $\eta$, and $\mathrm{Im}\,\omega$ all vanish, there is generically an $\mathcal O(1)$ phase contributing to baryogenesis coming from a combination of these individual phases. Indeed, even in the regions where the phases are aligned to give constructive or destructive interference in scattering rates (as discussed above), the combination of the LH neutrino phases $\delta+\eta$ is generally non-zero. For instance, with $\mathrm{Im}\,\omega\gg1$, constructive (destructive) interference in $\Gamma(L_e\rightarrow N^\dagger)$ occurs when $\delta+\eta$ is $\pi/2$ ($-\pi/2$), which is also the phase alignment that gives maximal $CP$ violation in the LH lepton sector. Therefore, $CP$ violation is generally $\mathcal{O}(1)$ over the entire parameter space of the minimal model. The vanishing of one phase (such as the experimentally accessible Dirac phase $\delta$) does not necessarily constrain the other phases relevant for leptogenesis \cite{Shaposhnikov:2008pf} or determine the relative lepton flavour scattering rates.

Finally, we note that there have been some  statements in the literature implying that the limit of large $|\mathrm{Im}\,\omega|$  somehow leads to more $CP$ violation. As shown in Eqs.~(\ref{eqn:tan_phi1})-(\ref{eqn:tan_phi2}), the $CP$ violating phases $\phi_1$ and $\phi_2$ depend on $\tanh(\mathrm{Im}\,\omega)$ and quickly saturate with increased $\mathrm{Im}\,\omega$. Instead, as discussed above, the main effect of $|\mathrm{Im}\,\omega|\gg1$ is an exponential enhancement of the Yukawa couplings, not the presence of an additional source of $CP$ violation.

%%%%%%%%%%%%%%%%
% Asymmetry evolution equations
%%%%%%%%%%%%%%%%

\section{Asymmetry Evolution Equations}
\label{sec:calculation}

In this section, we review the formalism for computing the baryon asymmetry from sterile neutrino oscillations. Along the way we discuss corrections we made to the existing formalism to account for the contribution of equilibrium processes to the evolution of the lepton asymmetries. Because of the central role of coherent oscillations among the sterile neutrino, the density matrix formalism is well-suited for tracking the evolution of abundances and coherences between states. Such an approach was used in \cite{Akhmedov:1998qx}, with  \cite{Asaka:2005pn} being the first analysis to include all of the  relevant terms in the evolution equations, with various factors corrected in subsequent work \cite{Asaka:2010kk,Asaka:2011wq}. More sophisticated approaches have also been taken, such as separately evolving different momentum modes in the density matrix \cite{Asaka:2011wq} or using non-equilibrium quantum field theory \cite{Drewes:2012ma}. These results are very similar to those using thermally averaged density matrix evolution; for example, ref.~\cite{Asaka:2011wq} find enhancements of the total baryon asymmetry of factors of 10-40\% when computing the asymmetries separately in each momentum mode relative to thermal averaging. Because of the computational simplicity of thermal averaging, and the small changes to the total asymmetry when using more sophisticated methods, we employ thermal averaging in the current work. 

As in earlier works, we follow the evolution of the sterile neutrino and anti-neutrino density matrices, $\rho_N$, and $\rho_{\bar N}$, as well as the LH lepton asymmetry density matrix, $\rho_{L-\bar{L}}$. Here, $L$ refers to a single component of the $\mathrm{SU}(2)$ lepton doublet, and is a matrix in lepton flavour space. The diagonal elements of the density matrices are equal to the  abundances of the corresponding fields normalized by the equilibrium abundance,
\be
\rho_{ii}(t) = \frac{Y_i(t)}{Y_i^{\rm eq}(t)},
\ee
and the off-diagonal terms correspond to the coherences between the fields. 

The evolution equations have the general  form  \cite{Asaka:2005pn},
\onecolumngrid 
\bea
\frac{d\rho_N}{dt} &=& -i [H(t),\,\rho_N] - \frac{1}{2}\{\Gamma(L^\dagger\rightarrow N)_{2\times2} ,\,\rho_N-\rho_{\bar L}^{\rm eq}\mathbb{I}_{2\times2}\} -\frac{1}{2}\gamma^{\rm av}T \,F^\dagger \rho_{L-\bar{L}}F,\label{eq:evol_first}\\
\frac{d\rho_{\bar N}}{dt} &=& -i [H(t),\,\rho_{\bar N}] - \frac{1}{2}\{\Gamma(L\rightarrow N^\dagger)_{2\times2} ,\,\rho_{\bar N}-\rho_L^{\rm eq}\mathbb{I}_{2\times2}\} + \frac{1}{2}\gamma^{\rm av}T\, F^{\rm T}\rho_{L-\bar{L}}F^*,\label{eq:evol_second}\\
\frac{d\rho_{L-\bar L}}{dt} &=& - \frac{1}{4}\{\Gamma(L\rightarrow N^\dagger)_{3\times3},\,\rho_{L-\bar L}\} +\frac{1}{2}\gamma^{\rm av}T\left( F\rho_{\bar N} F^\dagger-F^*\rho_{N} F^{\rm T}\right).\label{eq:evol_lepton}
\eea
\twocolumngrid
Here ($\{, \}$) $[ ,]$ denotes a matrix (anti-) commutator.  The evolution equations satisfy the relation $\mathrm{Tr}(\rho_{N-\bar N} - 2\rho_{L-\bar L}) = 0$, which reflects the conservation of the global $L-N$ charge. The interested reader can find a detailed account of these equations in Appendix~\ref{sec:fullevolution}. Here we instead concentrate on a qualitative understanding of the significance of the different terms that appear in Eqs.~(\ref{eq:evol_first})-(\ref{eq:evol_lepton}).
\benum
\item The Hamiltonian terms induce {\bf coherent oscillations} between diagonal and off-digonal components of the density matrix. Such a term is not included for the LH leptons because these rapidly decohere in the thermal bath of the early universe, and therefore only on-diagonal  density matrix components are non-zero.

\begin{figure}[t]
\centering
\includegraphics[width=0.25\textwidth]{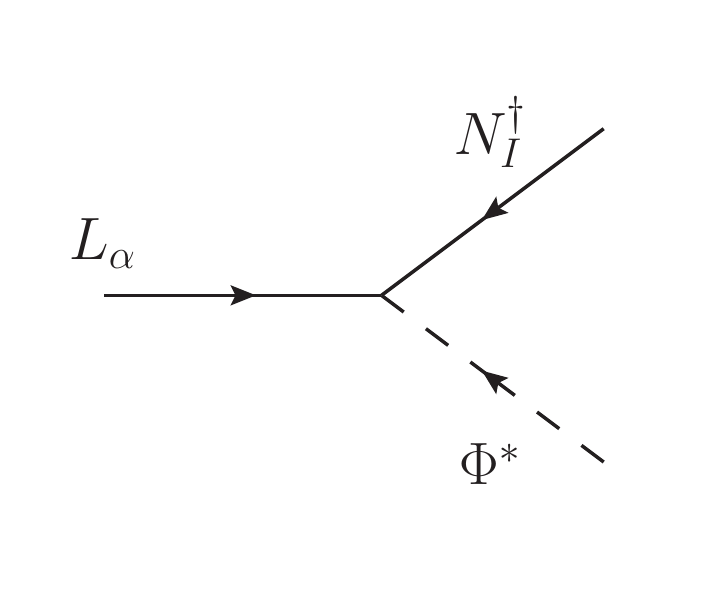}\hspace{0.7cm}\includegraphics[width=0.25\textwidth]{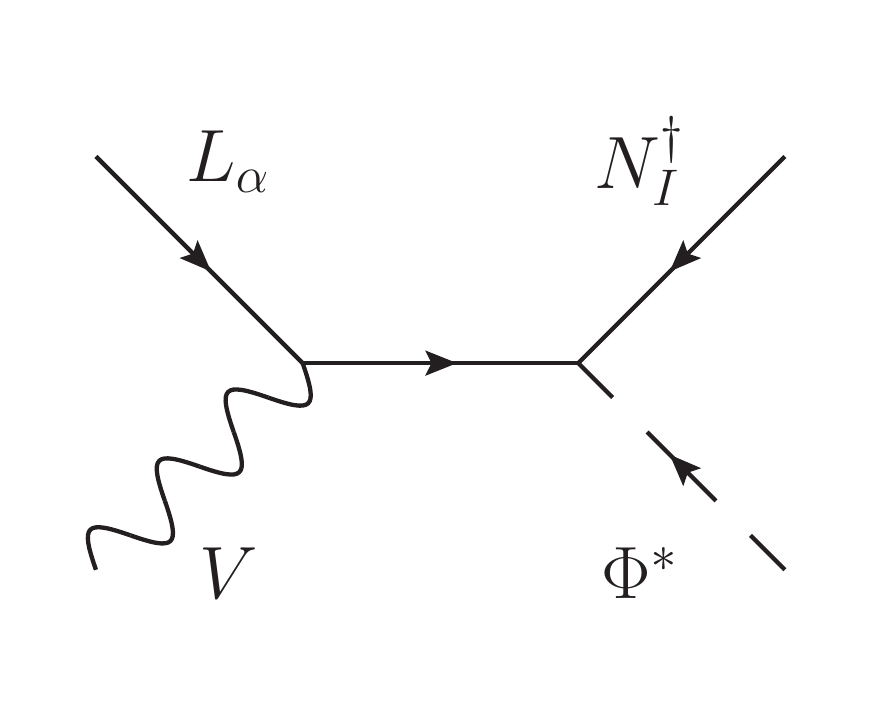}\hspace{0.7cm}\includegraphics[width=0.25\textwidth]{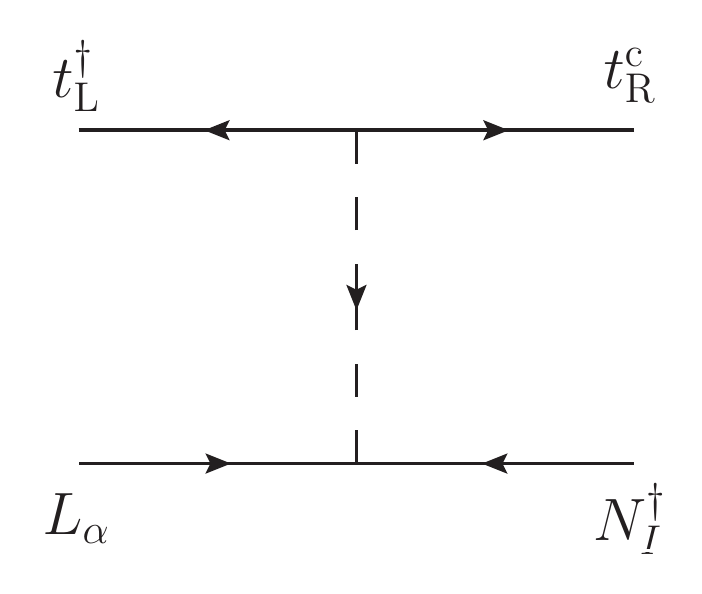}%
\caption{Representative Feynman diagrams for sterile neutrino creation  through (\textbf{Top}) $1\rightarrow2$ processes; (\textbf{Centre}) gauge boson $2\rightarrow2$ scattering; (\textbf{Bottom}) quark $2\rightarrow2$ scattering.}
\label{fig:Nfeynman}
\end{figure}

\item The terms proportional to $\Gamma(L\rightarrow N^\dagger)$ lead to both {\bf production} of the sterile neutrino abundance (through the terms $\propto \rho_L^{\rm eq}$) and {\bf destruction} (or washout) of the asymmetries in both $N$ and $L$. These interactions drive the fields towards their equilibrium distributions and erase any asymmetries. These scattering rates are proportional to the temperature $T$ and can often be factorized into the form
\be\label{eq:gamma_avg_def}
\Gamma(L\rightarrow N^\dagger) = \gamma^{\rm av}(T)\,T\,|F|^2.
\ee
The factor $\gamma^{\rm av}(T)$ parameterizes the rate after factoring out the Yukawa couplings and has been computed completely at leading order \cite{Besak:2012qm}. Representative diagrams contributing to the process are shown in Fig.~\ref{fig:Nfeynman}. The rates from ref.~\cite{Besak:2012qm} sum over both lepton doublet components.

\item The final terms in each line {\bf produce the asymmetries} in each sector. For $\rho_{L-\bar{L}}$, the coherent oscillations of the sterile neutrinos (encoded in the off-diagonal terms of $\rho_N$) create an asymmetry in individual LH lepton flavours at $\mathcal O(|F|^4)$. For the sterile neutrinos, an excess of $L_\alpha$ over $\bar L_\alpha$  translates into an excess of scattering into $N^\dagger$ vs.~$N$, sourcing an asymmetry in the sterile sector as well at $\mathcal O(|F|^6)$. The rate of transfer of SM lepton flavour asymmetries into sterile neutrino asymmetries depends on the asymmetry in the flavour $L_\alpha$ and the rate $L_\alpha \rightarrow N_I^\dagger$; the matrix $\rho_N$ appears between the $F^\dagger$ and $F$ matrices because of this sensitivity to lepton flavour effects.
\eenum
%

%%%%%%%%%%%%%%%% FIGURE %%%%%%%%%%%%%%%%%%%%%
%%%%%%%%%%%%%%%% FIGURE %%%%%%%%%%%%%%%%%%%%%

The initial conditions are such that there are initially no sterile neutrinos $\rho_N(0)=\rho_{\bar N}(0) =0$ and no primordial lepton asymmetry $\rho_{L-\bar L}(0)=0$. In existing works in the literature, these initial conditions are used to evolve Eqs.~(\ref{eq:evol_first})-(\ref{eq:evol_lepton}) down to the weak scale to determine the asymmetry $\rho_{L-\bar{L}}(t_{\rm W})$. At the weak scale, chemical potential relations relate the size of the baryon asymmetry to the total LH lepton asymmetry \cite{Harvey:1990qw}:
\be
\label{eq:YDeltaB_other_work}
Y_{\Delta B}(t_{\rm W}) = -\frac{28}{79}\sum_\alpha Y_{\Delta L_\alpha}(t_{\rm W}).
\ee
 Since sphalerons decouple at the electroweak phase transition, the final baryon asymmetry is frozen at this time and is simply $Y_{\Delta B}(t_{\rm W})$ as given in Eq.~(\ref{eq:YDeltaB_other_work}). 
 
 The above approach to solving for the late-time baryon asymmetry is  incorrect, as it neglects the effects of interactions between SM fields, which are rapid compared to the active-sterile neutrino scatterings. Because of these  scatterings, the asymmetries in individual lepton flavours  created by sterile neutrino oscillations are rapidly distributed among all SM fields. Since the asymmetries are destroyed only through interactions of the LH leptons with sterile neutrinos, this modifies the relative rates of asymmetry creation and destruction. 
 
The effects of equilibrium scatterings on the evolution of an asymmetry have been understood and corrected in a different context, namely of baryogenesis through weak-scale dark matter scatterings \cite{Cui:2011ab}. Our approach here is similar: we include in our density matrices only quantities that are preserved by the equilibrium SM interactions, justifying the absence of such rapid interactions in the evolution equations. Specifically, we exchange the anomalous asymmetries in individual lepton flavours $L_\alpha$ in (\ref{eq:evol_lepton}) for asymmetries of $B-3L_\alpha$, which are exactly conserved by SM scatterings. This modifies the density matrix equations: because Eq.~(\ref{eq:evol_lepton}) now represents the evolution of $\Delta(B-3L_\alpha)$, instead of $\Delta L_\alpha$, all terms are a factor of $-2\times3$ larger than for the individual LH lepton species, due to the re-definition of the charge and an $\mathrm{SU}(2)$  factor from summing over charged and neutral components of $L_\alpha$. The last term in Eq.~(\ref{eq:evol_lepton}), which creates the individual $\Delta L_\alpha$ asymmetries, is otherwise unmodified, since the coherent scattering $N^\dagger\rightarrow L_\alpha$ generating the asymmetry is not sensitive to the details of SM thermal scatterings. By contrast, the first term in Eq.~(\ref{eq:evol_lepton}), which destroys the individual lepton flavour asymmetries, is proportional \emph{only} to the individual asymmetry $\Delta L_\alpha$, \emph{not} the total $\Delta(B-3L_\alpha)$ asymmetry\footnote{The same is true for the terms in the kinetic equations generating the sterile neutrino asymmetries.}. This is because the asymmetry in fields carrying $B-3L_\alpha$ is divided by the equilibrium scatterings among many SM fields, such as the quarks and RH charged leptons, which have no direct coupling to sterile neutrinos. To write the correct evolution equation, we relate the LH lepton abundances from Eq.~(\ref{eq:evol_lepton}) to the $B-3L_\alpha$ charges through the chemical potential constraints of the SM interactions:
\be
\nonumber
(\rho_{L-\bar L})_\alpha = -\frac{1}{2133}\sum_\beta\left[221\,\delta_{\alpha\beta} -16(1-\delta_{\alpha\beta})\right]\,\rho_{B-3L_\alpha}.
\\ \label{eq:chem_relations}~
\ee
We see immediately that the rates are suppressed by at least $\approx 6\times221/2133\approx0.6$ due to the fact that much of the asymmetry leaks out of the LH leptons, and possibly more due to the other negative terms in the rotation matrix. Having made this substitution, we  express the density matrix evolution equations purely in terms of $\rho_N$, $\rho_{\bar N}$, and $\rho_{B-3L_\alpha}$. These quantities are then evolved down to the weak scale, and instead of Eq.~(\ref{eq:YDeltaB_other_work}) the final baryon asymmetry is
\be
Y_{\Delta B}(t_{\rm W}) = \frac{28}{237}\sum_\alpha Y_{\Delta(B-3L_\alpha)}(t_{\rm W}).
\ee
The equations are actually easiest to solve in a  different basis: we provide the complete evolution equations in this basis and a few details on numerical integration in Appendix \ref{sec:fullevolution}. In that same appendix, we also include the effects arising from a phase-space suppression of certain diagrams contributing to the $\Delta L_\alpha$ destruction rates; such modifications were first discussed in \cite{Asaka:2011wq}.

%To summarize the effects of equilibrium scatterings on the density matrix evolution equations, we re-write them in the form used in our updated analysis:
%
%\bea
%\frac{d\rho_N}{dt} &=& -i [H(t),\,\rho_N] - 2\{\Gamma(L\rightarrow N^\dagger) ,\,\rho_N-\rho_{\bar L}^{\rm eq}\mathbb{I}_{2\times2}\} -\frac{1}{2} \gamma^{\rm av}T \,F^\dagger \rho_{L-\bar L}F,\label{eq:evol_first}\\
%\frac{d\rho_{\bar N}}{dt} &=& -i [H(t),\,\rho_{\bar N}] - 2\{\Gamma(L\rightarrow N^\dagger) ,\,\rho_{\bar N}-\rho_L^{\rm eq}\mathbb{I}_{2\times2}\} + \frac{1}{2}\gamma^{\rm av}T\, F^{\rm T} \rho_{L-\bar L}F^*,\label{eq:evol_second}\\
%\frac{d\rho_{\Delta(B-3L_\alpha)}}{dt} &=& 6\{\Gamma(L\rightarrow N^\dagger),\,\rho_{L-\bar L}\} -6\gamma^{\rm av}T\left( F\rho_{\bar N} F^\dagger-F^*\rho_{N} F^{\rm T}\right),\label{eq:evol_lepton}
%\eea
%
%with $\rho_{L-\bar L}$ is substituted with the value (\ref{eq:chem_relations}). The equations are actually easiest to solve in a slightly different basis: we provide the complete evolution equations and a few details on numerical integration in Appendix \ref{sec:fullevolution}.

%%%%%%%%%%%%%%%%
% BAU Minimal Model
%%%%%%%%%%%%%%%%
\section{Baryon Asymmetry in the $\nu$MSM}
\label{sec:BAU_minimal}

This section is devoted to a quantitative confirmation of the qualitative features discussed above. We demonstrate the continuous transition between the different parameter regimes discussed in section~\ref{sec:mechanism} and depicted in Fig.~\ref{fig:asymmetry_regimes}. We also show that, over the entire parameter space of the minimal model, the observed baryon asymmetry requires a very specific alignment of the model parameters to within at least $\mathcal{O}(10^{-5})$. We do not perform a comprehensive scan of  possible parameters in the $\nu$MSM, as such a scan was carried out in \cite{Canetti:2010aw,Canetti:2012kh} for Regimes I and II with kinetic equations similar to those we use here, and our results are in qualitative agreement with theirs. However,  our qualitative picture provides a simple understanding for the parameters maximizing the baryon asymmetry found in ref.~\cite{Canetti:2010aw,Canetti:2012kh}. Finally, with regime III we explore a new region of parameter space whose existence was first demonstrated in ref.~\cite{Drewes:2012ma} with a couple of representative points in a model with three sterile neutrinos (instead of the two we consider here). We expand upon ref.~\cite{Drewes:2012ma} with a comprehensive scan of the parameter space and demonstrate the emergence of Regime III as a continuous part of the other regimes. 

The observed baryon asymmetry of the Universe is \cite{Ade:2013zuv}
\be
Y_{\Delta B} \approx 8.6\times 10^{-11}.
\ee
Throughout this section,  we take $T_{\rm W}=140$ GeV as the temperature of the electroweak phase transition, corresponding to a SM Higgs mass of 126 GeV \cite{Canetti:2012kh}. For the numerical solutions of the evolution equations in all regimes, we take values consistent with  Section \ref{sec:models}: $m_1=0$, $m_2=9$ meV, $m_3=49$ meV, $\sin\theta_{12}=0.55$, $\sin\theta_{23}=0.63$, and $\sin\theta_{13}=0.16$. For concreteness, we use $\mathrm{Re}\,\omega=\pi/4$ throughout our analyses, as this gives an appreciable asymmetry for all values of $\mathrm{Im}\,\omega$. The asymmetry does not change substantially with this angle: for example, $\mathrm{Re}\,\omega=\pi/2$ gives a comparable baryon asymmetry, with somewhat larger values than $\mathrm{Re}\,\omega=\pi/4$ for $\mathrm{Im}\,\omega\approx0.5-3$, and smaller values elsewhere (see Fig.~\ref{fig:tuning_measure}).
%
%Following our intuition from Section \ref{sec:mechanism}, the generated asymmetry in the $\nu$MSM model is largest when:
%%
%\benum
%\item The sterile neutrinos are degenerate ($\Delta M_N/M_N$), delaying oscillations until the sterile neutrino abundance is larger. 
%\item The Yukawa couplings are as large as possible without washing out all lepton flavour asymmetries or causing decoherence of the neutrino oscillations before an asymmetry is created. In the minimal model, the Yukawa couplings can be tuned large by increasing the complex mixing angle $|\mathrm{Im}\,\omega|$. 
%\item There are large differences in the scattering rates of individual lepton flavours into sterile neutrinos. These differences arise due to either hierarchical parameters in the neutrino mixing matrix (light neutrino masses or $\theta_{13}\ll\theta_{12},\theta_{23}$) or through strong constructive/destructive interference in particular flavour rates induced by alignments of the $CP$ phases. 
%
%\eenum
%%
%Earlier studies have focused predominantly on scenarios where highly degenerate sterile-neutrino spectra are relied upon to obtain the baryon asymmetry of the universe, although \cite{Drewes:2012ma} showed that this condition was unnecessary with three sterile neutrinos participating in leptogenesis.
%
%We illustrate the importance of  these factors for the three regimes discussed in Section \ref{sec:mechanism} and Fig.~\ref{fig:asymmetry_regimes}. 

\subsection{Regime I}
The generation of a baryon asymmetry in this regime is maximized when the rates of LH lepton scattering into sterile neutrinos are very different for different lepton flavours. However, the Yukawa couplings here are set to their small, natural values expected from the see-saw relation, Eq.~(\ref{eq:seesaw_lag}). Therefore, even with large differences in lepton flavour scattering rates, the baryon asymmetry in Regime I is not large enough to account for the observed asymmetry unless the coherent oscillation time is maximal, which requires a strong degeneracy between the masses of $N_2$ and $N_3$. A mass degeneracy of $\Delta M_N/M_N\sim \mathcal{O}(10^{-6}-10^{-8})$ is needed, depending on $\mathrm{Im}\,\omega$. 

\begin{figure}[t]
\centering
\includegraphics[width=0.45\textwidth]{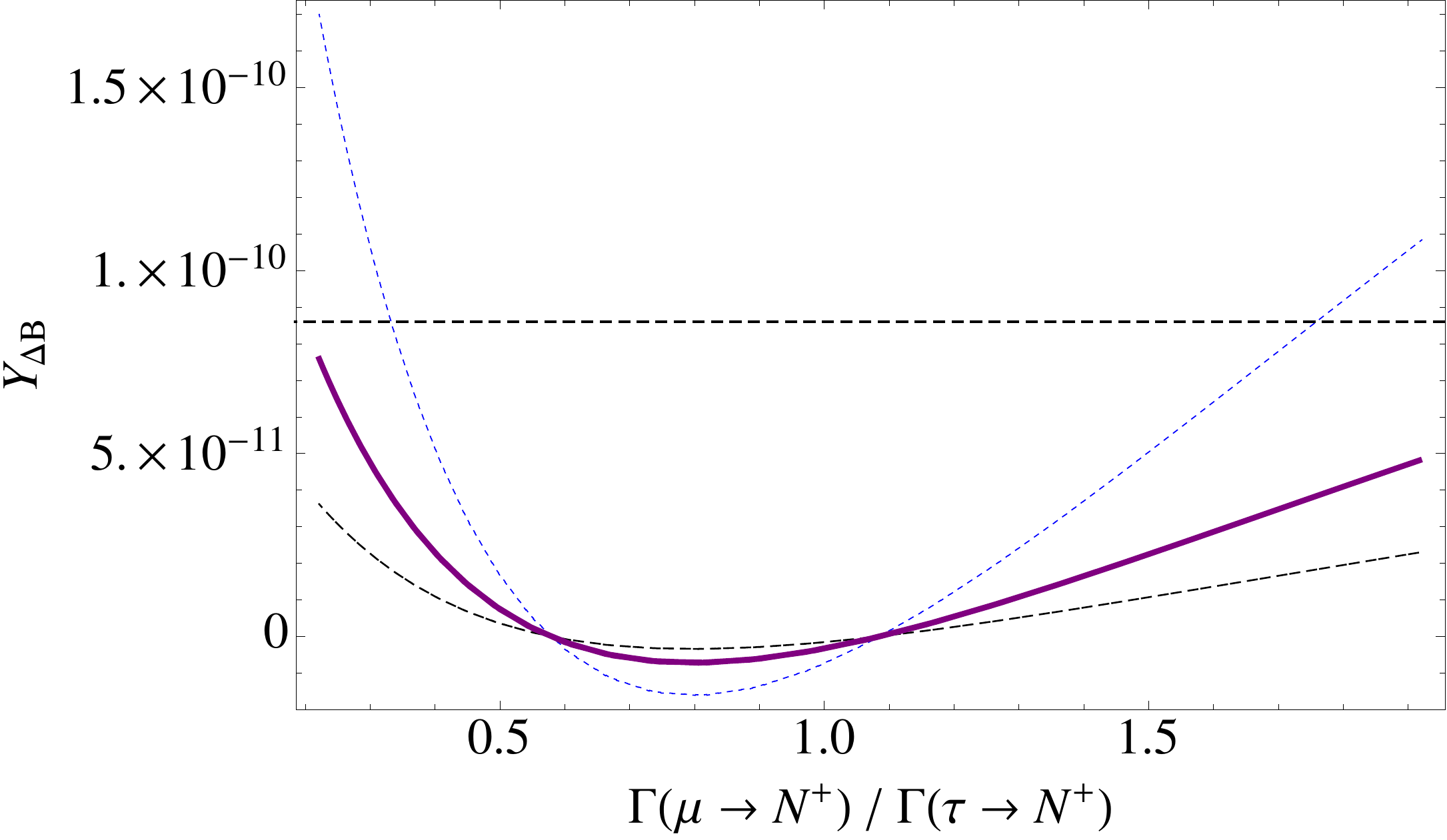}%
\caption{Illustration of the importance of flavour-dependent effects in generating a non-zero baryon asymmetry in Regime I. The baryon asymmetry is plotted as a function of the ratio between the muon-sterile neutrino scattering rate and the corresponding tau-sterile rate. The asymmetry is maximized when the washout rates for $\mu$ and $\tau$ are substantially different, as predicted by Eq.~(\ref{eq:totalbaryon}). The mass splittings are $\Delta M_N =3\times10^{-8}$ GeV (blue, short dash), $10^{-7}$ GeV (purple, solid), $3\times10^{-7}$ GeV (black, long dash). The ratio of muon to tau rates is varied by changing the relative values of $\delta-\eta$, while the overall MNS $CP$ phase and other parameters are held fixed ($M_N=1$ GeV, $\omega = \pi/4 - i/2$, $\delta+\eta=\pi/2$). The horizontal dashed line indicates the observed baryon asymmetry.}
\label{fig:regimeIplot}
\end{figure}

To illustrate this, we show in Fig.~\ref{fig:regimeIplot} the baryon asymmetry for a set of parameters where the magnitude of the Yukawa coupling $F$ is held fixed, while the relative rates of $L_e\rightarrow N^\dagger$, $L_\mu\rightarrow N^\dagger$, and $L_\tau\rightarrow N^\dagger$ are allowed to vary. We see that the asymmetry is largest when flavour dependence on the scattering rates is significant, i.e.~the rates of $\Gamma(L_\alpha\rightarrow N^\dagger)$ are very different, especially the rates of $L_\mu$ and $L_\tau$. In this regime, the contribution from the $L_e$ flavour asymmetry and scattering rate is subdominant, due to the smallness of  $\theta_{13}$ and the fact that $m_3$ has the largest Yukawa coupling to the sterile sector. Over the parameters scanned in Fig.~\ref{fig:regimeIplot}, the rate $\Gamma(L_e\rightarrow N^\dagger)$ is typically $\lesssim3\%$ of the corresponding $L_\mu$ and $L_\tau$, and the asymmetry in $L_e$ is $\lesssim10\%$ of the $L_\mu$ and $L_\tau$ asymmetries\footnote{This is true except in a small window around $\Gamma(L_\mu\rightarrow N^\dagger)/\Gamma(L_\tau\rightarrow N^\dagger)\approx0.7$, where the $L_\tau$ asymmetry becomes very small and the $L_e$ asymmetry is important.}.   The asymmetry vanishes when $\Gamma(L_\mu\rightarrow N^\dagger)/\Gamma(L_\tau\rightarrow N^\dagger)\approx0.8$, which is when the difference in $L_\mu\rightarrow N^\dagger$ and $L_\tau\rightarrow N^\dagger$ rates exactly compensates for the difference in flavour asymmetries.

%%%%%%%%%%%%%%% FIGURE %%%%%%%%%%%%%%%%%%%%%
%%%%%%%%%%%%%%% FIGURE %%%%%%%%%%%%%%%%%%%%%

This behaviour is easily understood with the qualitative lessons learned in previous sections. In this regime, the Yukawa couplings are sufficiently small that equilibration of the active and sterile neutrinos is irrelevant as it occurs long after the electroweak time scale. As a result, the asymmetries in individual lepton flavours remain unchanged from their generation at the time of coherent sterile neutrino oscillation to the sphaleron decoupling at $T_{\rm W}$. The baryon asymmetry is determined by the slow transfer of asymmetry from individual lepton flavours into the sterile sector, Eq.~(\ref{eq:totalbaryon}), which generates a total lepton asymmetry. It is therefore maximized with large differences in rates associated with different lepton flavours.

\subsection{Regime II} 
As discussed in Section~\ref{sec:mechanism}, the maximization of the baryon asymmetry in this region is done by increasing the scattering rates as much as possible while avoiding equilibration (and subsequent washout) before the electroweak time scale,  $\Gamma(L_\alpha\rightarrow N^\dagger)\sim H$ when $T\sim T_{\rm W}$. This optimization is achieved by setting the Yukawa couplings to be much larger than their natural see-saw relations through a careful alignment of the different Yukawa couplings so that $|F|^2/F^2 \sim \cosh(2{\rm Im}\omega)  \sim \mathcal{O}(10^2)$.  
 
%%%%%%%%%%%%%%% FIGURE %%%%%%%%%%%%%%%%%%%%%
\begin{figure}[t]
\centering
\includegraphics[width=0.5\textwidth]{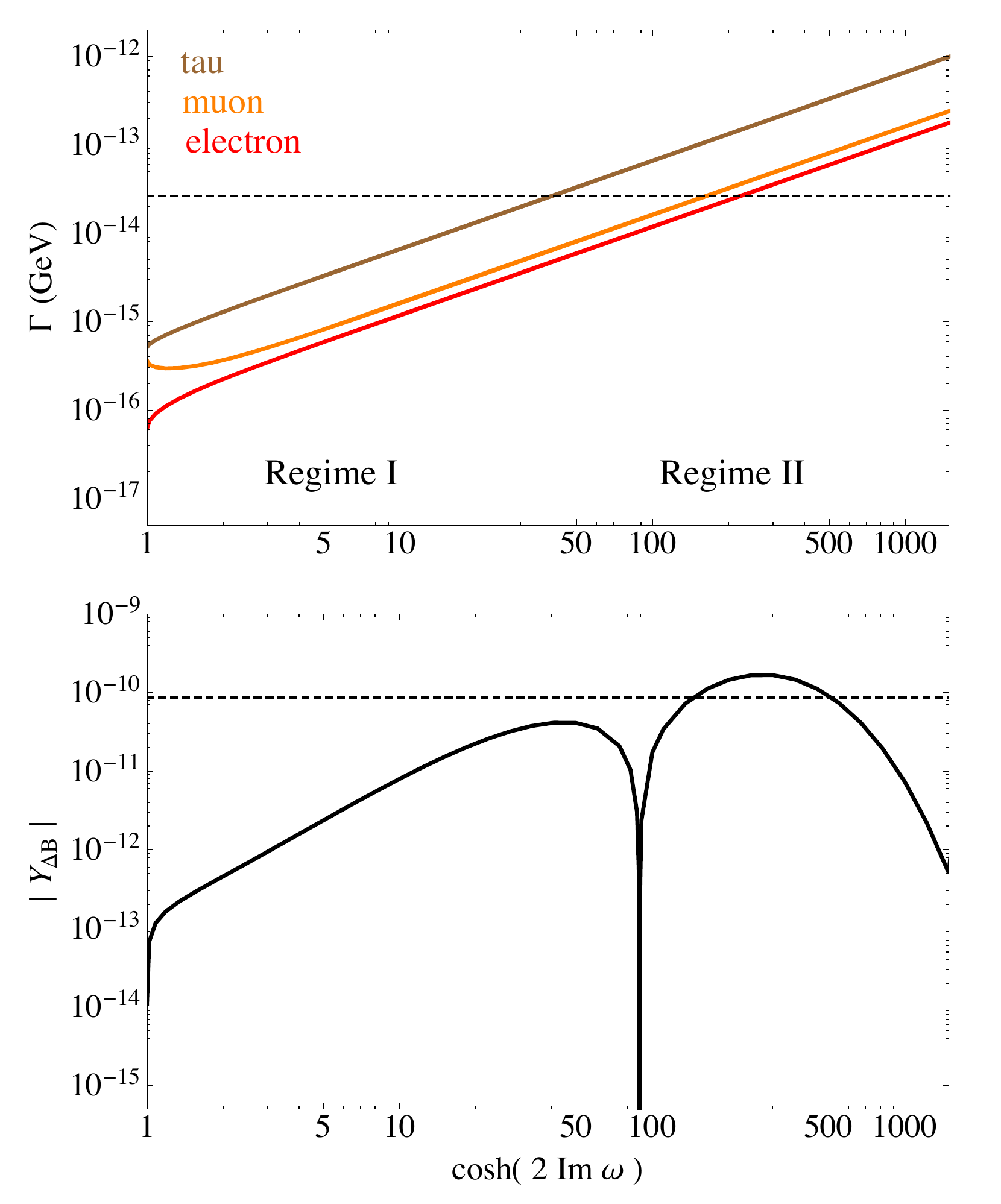}%
\caption{Baryon asymmetry (\textbf{Bottom}) and lepton washout rates at $T=140$ GeV (\textbf{Top}) as a function of the Yukawa coupling magnitude, which is parameterized by $\mathrm{Im}\,\omega$. The plot shows the continuous evolution from Regime I to Regime II. The washout rates are shown for taus (top), muons, and electrons (bottom); the dashed line in the upper plot shows the Hubble scale at $T_{\rm W}=140$ GeV, while the dashed line in the lower plot shows the observed $Y_{\Delta B}=8.6\times10^{-11}$. The asymmetry is maximized  around $\Gamma(L_e\rightarrow N^\dagger)\sim\Gamma(L_\mu\rightarrow N^\dagger) \approx H$. Other parameters held fixed: $M_N=1$ GeV, $\Delta M_N = 10^{-5}$ GeV, $\eta=-\pi/4$, $\delta=3\pi/4$, $\mathrm{Im}\,\omega>0$.}
\label{fig:regimeIIplot_constr}
\end{figure}
%%%%%%%%%%%%%%% FIGURE %%%%%%%%%%%%%%%%%%%%%

To illustrate this, we show in Fig.~\ref{fig:regimeIIplot_constr} the magnitude of the baryon asymmetry as a function of Yukawa coupling magnitude. We set the Yukawa coupling magnitude by changing $\mathrm{Im}\,\omega$. Unlike in Regime I, the $L_e$ flavour here is important; although the flavour asymmetry in $L_e$ is typically smaller than in $L_\mu$ or $L_\tau$, it also equilibrates more slowly, and so can be comparable to the asymmetries in $L_\mu$ or $L_\tau$ if the latter are partially washed out. As expected, the magnitude of the baryon asymmetry increases monotonically with the Yukawa coupling, except for a small region where the  asymmetry changes sign due to the onset of $L_\tau\rightarrow N^\dagger$ equilibration, which modifies the total lepton asymmetry as discussed in Section \ref{sec:mechanism}. The total baryon asymmetry is maximized in the region with $\Gamma(L_\alpha\rightarrow N^\dagger)\sim H(T_{\rm W})$. Any further increase beyond this point results in an equilibration time earlier than the electroweak scale and a precipitous decrease in the baryon asymmetry.

For the parameter points in Fig.~\ref{fig:regimeIIplot_constr}, we see that the observed baryon asymmetry is obtained in Regime II with $\Delta M_N / M_N\sim 10^{-5}$, which is less degenerate than in Regime I. However, this reduction in parameter tuning from mass degeneracy is compensated by a tuning of the Yukawa couplings that goes like one part in $\cosh2\,\mathrm{Im}\,\omega\sim\mathcal{O}(10^2)$ in Regime II. Therefore, the parameter space giving the observed baryon asymmetry is no less tuned than in Regime I, but the tuning arises from alignments in both the sterile neutrino masses and the Yukawa couplings.

\subsection{Regime III}
In Regime III, the effects of large Yukawa couplings and maximal differences in active-sterile neutrino scattering rates combine to give the largest possible asymmetry. The Yukawa couplings are set even larger than in Regime II, enhancing the generated baryon asymmetry. As expected, such large Yukawas also cause the equilibration time scale to occur earlier than the electroweak scale. If this were true for all lepton flavours, it would have resulted in complete washout of the asymmetry, which is phenomenologically unacceptable. However, in this part of parameter space, destructive interference in the $L_e\rightarrow N^\dagger$ rate allows the $L_e$ asymmetry to avoid early washout by remaining out of equilibrium until the electroweak scale when $\Gamma(L_e\rightarrow N^\dagger)\approx H(T_{\rm W})$. This effect was first discussed in the context of leptogenesis with three sterile neutrinos \cite{Drewes:2012ma}.

A strong destructive interference in the rate $\Gamma(L_e\rightarrow N^\dagger)$ is possible when $\mathrm{Im}\,\omega\gg1$ and \cite{Asaka:2011pb,Drewes:2012ma}
\bea
\tan\theta_{13} &=& \frac{m_2}{m_3}\sin\theta_{12}, \label{eq:angle_alignment}\\
\cos(\delta+\eta) &=& -1. \label{eq:phase_alignment}
\eea
The parameters in Eq.~(\ref{eq:angle_alignment}) are fixed by oscillation data. Interestingly, the current best-fit value of $\theta_{13}$ happens to be very close to satisfying Eq.~(\ref{eq:angle_alignment}), leading to very strong suppression of $\Gamma(L_e\rightarrow N^\dagger)$ when the Majorana and Dirac phases satisfy Eq.~(\ref{eq:phase_alignment}). In practice, the destructive interference is still very effective if $\delta+\eta$ are within about $10\%$ of this critical value, but cannot deviate much more than this. The $L_e$ asymmetry generation rate is proportional to $\sin(2\mathrm{Re}\,\omega)$ in the destructive interference limit $\cos(\delta+\eta)\rightarrow0$, and so the baryon asymmetry is maximized for $\mathrm{Re}\,\omega\approx\pi/4$.

\begin{figure}[t]
\centering
\includegraphics[width=0.5\textwidth]{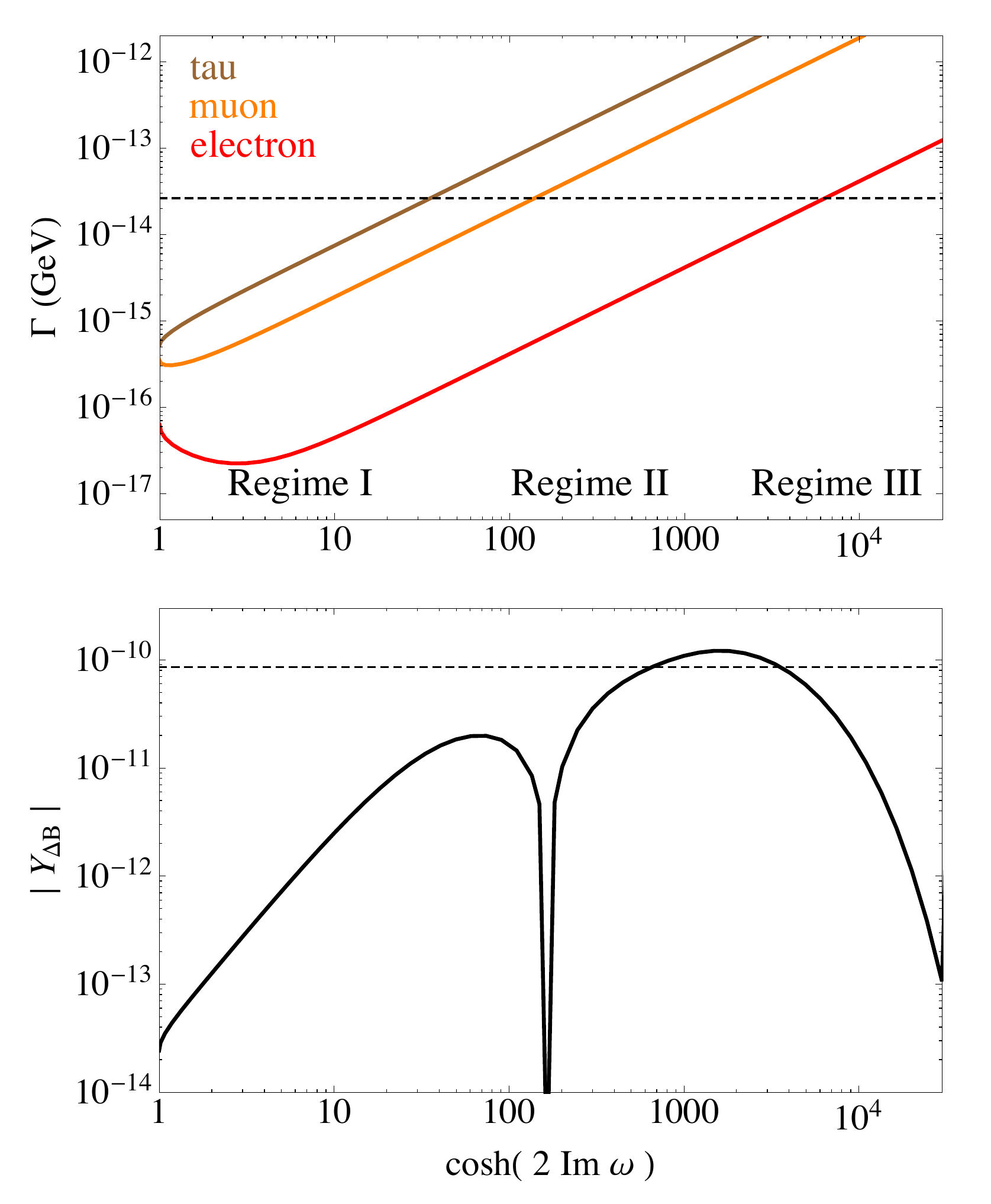}%
\caption{Baryon asymmetry (\textbf{Bottom}) and lepton washout rates at $T=140$ GeV (\textbf{Top}) as a function of the Yukawa coupling magnitude, which is parameterized by $\mathrm{Im}\,\omega$. The plot shows the continuous evolution from Regime I to Regime III for parameters exhibiting strong destructive interference in $\Gamma(L_e\rightarrow N^\dagger)$. The curves are the same as Fig.~\ref{fig:regimeIIplot_constr}. The asymmetry is again maximized around $\Gamma(L_e\rightarrow N^\dagger) \approx H$. Other parameters held fixed: $M_N=1$ GeV, $\Delta M_N = 10^{-3}$ GeV, $\eta=\delta=-\pi/4$, $\mathrm{Im}\,\omega>0$.}
\label{fig:regimeIIplot_destr}
\end{figure}

In Fig.~\ref{fig:regimeIIplot_destr}, we show the baryon asymmetry for a choice of parameters satisfying (\ref{eq:phase_alignment}). The Yukawa couplings are changed by varying $\mathrm{Im}\,\omega$, and we demonstrate how the asymmetry varies continuously from Regimes I to III. As in Fig.~\ref{fig:regimeIIplot_constr}, the magnitude of the baryon asymmetry monotonically increases (except when it changes sign at the point of $L_\tau\rightarrow N$ equilibration). Once again, the baryon asymmetry is maximal when the Yukawa couplings have a value such that $L_e$ is just coming into equilibrium at $T_{\rm W}$, while $L_\mu$ and $L_\tau$ equilibrate at earlier times. 

%%%%%%%%%%%%%%% FIGURE %%%%%%%%%%%%%%%%%%%%%
%%%%%%%%%%%%%%% FIGURE %%%%%%%%%%%%%%%%%%%%%

As a result of the enhanced asymmetry in Regime III, the required mass degeneracy for the baryon asymmetry of the universe is the smallest of any region of parameter space. However, even here a degeneracy of $\Delta M_N/M_N \sim 10^{-3}$ is necessary, as is a tuning of the Yukawa coupling of one part in $10^4$ and an alignment of the $CP$ phases $\delta+\eta\approx-\pi/2$ to within 10\%. As in all of the other regimes, the combined tuning in mass degeneracy and Yukawa couplings is larger than one part in $10^5$.

\subsection{Tuning}
\label{sec:tuning}

\begin{figure}[t]
\centering
\includegraphics[width=0.4\textwidth]{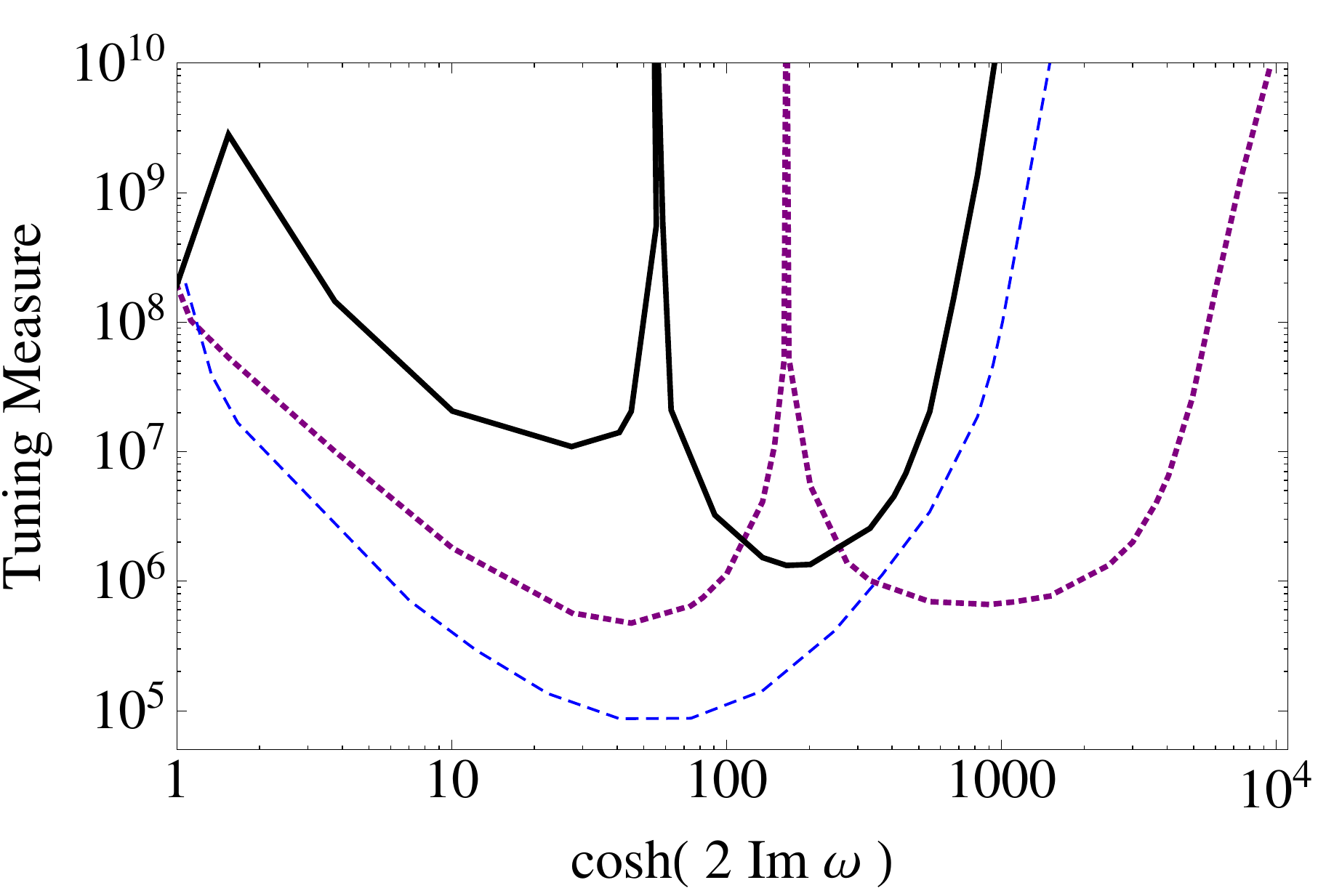}%
\caption{Fine-tuning required to obtain the observed baryon asymmetry, $Y_{\Delta B}=8.6\times10^{-11}$. The tuning measure is $\cosh(2\mathrm{Im}\,\omega)\,M_N/\Delta M_N$. The other parameters used are (black, solid) $\eta =\delta=\mathrm{Re}\,\omega=\pi/4$; (purple, dotted) $\eta=\delta=-\mathrm{Re}\,\omega=-\pi/4$; (blue, dashed) $\eta=2.42$, $\delta=0.5$, $\mathrm{Re}\,\omega=\pi/2$. For all points, $\Delta M_N$ is fixed by imposing the observed baryon asymmetry.}
\label{fig:tuning_measure}
\end{figure}

To close this section, we show in Fig.~\ref{fig:tuning_measure}  the  tuning necessary in the different regimes in one continuous plot. We take the tuning to be the product of the tuning of the mass splitting, $M_N/\Delta M_N$, and the alignment of the Yukawa couplings, $\cosh(2{\rm Im}\,\omega)$. While such tunings/alignments are technically natural, there is no explanation for their structure in the minimal model. This might simply be a feature of nature, but it could also be a hint for  additional structure beyond the $\nu$MSM. The case for considering an extended model becomes particularly cogent if it brings with it new observable effects. We present one such extension in the following section, show that it entirely alleviates the needed tuning, and discuss its observability in on-going collider searches. 

%%%%%%%%%%%%%%% FIGURE %%%%%%%%%%%%%%%%%%%%%
%%%%%%%%%%%%%%% FIGURE %%%%%%%%%%%%%%%%%%%%%

%%%%%%%%%%%%%%%%
% BAU Extended Models
%%%%%%%%%%%%%%%%

\section{Baryon Asymmetry with a Leptophilic Higgs}
\label{sec:BAU_extended}

In the previous section, we saw that increasing the Yukawa couplings can generally lead to larger asymmetry generation. Schematically, we have that~\cite{Asaka:2005pn,Shaposhnikov:2008pf,Asaka:2010kk}, 
\bea
\nonumber
\begin{array}{c}
{\rm asymmetry}   \\
{\rm generation~rate}     
\end{array}
&\propto& \mathrm{Im}(FF^\dagger)_{\alpha\alpha}^2 \sim \frac{m_\nu^2 M_N^2}{\langle\Phi\rangle^4}\cosh(2\mathrm{Im}\,\omega).
\\&~&\label{eq:generationrate}
\eea
For very small Yukawas (regime I), the asymmetry generation rate is correspondingly small and one must tune the mass splitting of the sterile neutrino so that the oscillation time scale is as long as possible and approaches the electroweak time scale. We saw that it is possible to increase the asymmetry generation rate, and thereby alleviate some of this tuning,  by increasing the Yukawa couplings (while keeping the seesaw relations intact) with an alignment controlled by the parameter ${\rm Im}\,\omega$ (regime II). However, this increase cannot proceed indefinitely since, at some point, the equilibration time scale becomes as early as the electroweak time scale. At this point, the washout processes that kick-in and act to reduce the asymmetry scale as
\bea
\nonumber
\begin{array}{c}
{\rm asymmetry}   \\
{\rm washout~rate}     
\end{array}
  &\propto& (FF^\dagger)_{\alpha\alpha} \sim \frac{m_\nu M_N}{\langle\Phi\rangle^2}\cosh(2\mathrm{Im}\,\omega).
\\&~& \label{eq:washoutrate}
\eea
The scaling relations Eqs.~(\ref{eq:generationrate}) and~(\ref{eq:washoutrate}) show that, while one can initially enhance the asymmetry by increasing the alignment in the Yukawas, $\cosh(2\mathrm{Im}\,\omega) \gg 1$, this gain is saturated once the time scale of equilibration coincides with the electroweak scale and washout processes become relevant. Regime III circumvents this saturation and supports even stronger alignment in the Yukawas by having a different equilibration time scale for the different lepton flavours. 

However, the above scaling relations suggest an alternative approach. The asymmetry generation rate, Eq.~(\ref{eq:generationrate}), depends strongly on the value of the electroweak VEV, $\langle\Phi\rangle$. It can be greatly enhanced if the LH neutrino masses arise from a new source of electroweak symmetry breaking. Such a scenario can arise in a leptophilic Two Higgs Doublet Model (2HDM), with one scalar coupling exclusively to leptons and acquiring a much smaller VEV. This idea was also mentioned in ref.~\cite{Drewes:2012ma} as a way to alleviate the needed alignment in the Yukawa couplings. Here, however, we see that a smaller leptophilic Higgs VEV has a much more pronounced effect than just tuning the Yukawa couplings to be large in the minimal model ($|\mathrm{Im}\,\omega|\gg1$):  the asymmetry generation and washout rates scale very differently with the VEV, while they have the same scaling with $\mathrm{Im}\,\omega$.

\subsection{The Model}
A Z$_2$ symmetry is typically required to prevent both Higgs fields from coupling to the same fermions, inducing tree-level flavour-changing neutral currents \cite{Glashow:1976nt}. One possible choice is the {\bf leptophilic} (or {\bf Type IV}) 2HDM:
\bea
\mathcal L_{\mathrm{leptophilic}} &=& \mu_1^2|\Phi_1|^2 - \mu_2^2|\Phi_2|^2  - \frac{\lambda_1}{4}|\Phi_1|^4 -\frac{\lambda_2}{4}|\Phi_2|^4 \nonumber
\\\nonumber
&& {} + \lambda_u\, Q \Phi_1 u^{\rm c} + 
\lambda_d \,Q \Phi_1^* d^{\rm c} 
\\
&& {} + \lambda_\ell\, L \Phi_2^* E^{\rm c} + F\, L \Phi_2 N + \mathrm{h.c.}
\eea
In this model, $\Phi_1$ is a SM-like Higgs giving mass to the quarks, and $\Phi_2$ is a leptophilic Higgs giving mass to the charged leptons and LH neutrinos. It is motivated by the observation that the heaviest lepton masses are much smaller than the heaviest quark masses, and  might obtain their masses through a field with a smaller VEV. Indeed,  $\lambda_\tau\sim\mathcal O(1)$ for $\langle\Phi_2\rangle \approx2$ GeV.

If $\Phi_2$ acquires a VEV through a negative mass-squared term in the potential, a prediction of the associated scalar masses would be $m^2 \sim \lambda \langle\Phi_\ell\rangle^2$. Collider searches  rule out the existence of any such charged states below 100 GeV. Therefore, $\Phi_2$ must instead acquire a VEV through a linear tadpole term in its potential. We assume this comes from a mixing with the SM Higgs, which arises from additional terms in the 2HDM potential:
\be\label{eq:2HDM_potential}
V_{\rm 2HDM} \supset  \mu_{\rm mix}^2 \Phi_{1}\Phi^*_2 + \mathrm{h.c.}
\ee
$\mu_{\rm mix}^2$ can be naturally smaller than $\mu_1^2$ and $\mu_2^2$ since it  breaks the Z$_2$ symmetry\footnote{For our purposes, other Z$_2$-breaking terms such as $|\Phi_1|^2 \Phi_1\Phi_2^*$ can be absorbed  into a redefinition of $\mu_{\rm mix}$.}. The ratio of VEVs is
\be\label{eq:tanbeta}
\tan\beta \equiv \frac{\langle\Phi_1\rangle}{\langle\Phi_2\rangle}\approx  \frac{\mu_2^2}{\mu_{\rm mix}^2},
\ee
which can account for the observed pattern $\langle\Phi_1\rangle\gg \langle\Phi_2\rangle$ for $\mu_2\gg \mu_{\rm mix}$. 

\subsection{Asymmetry Generation}

\begin{figure}[t]
\centering
\includegraphics[width=0.44\textwidth]{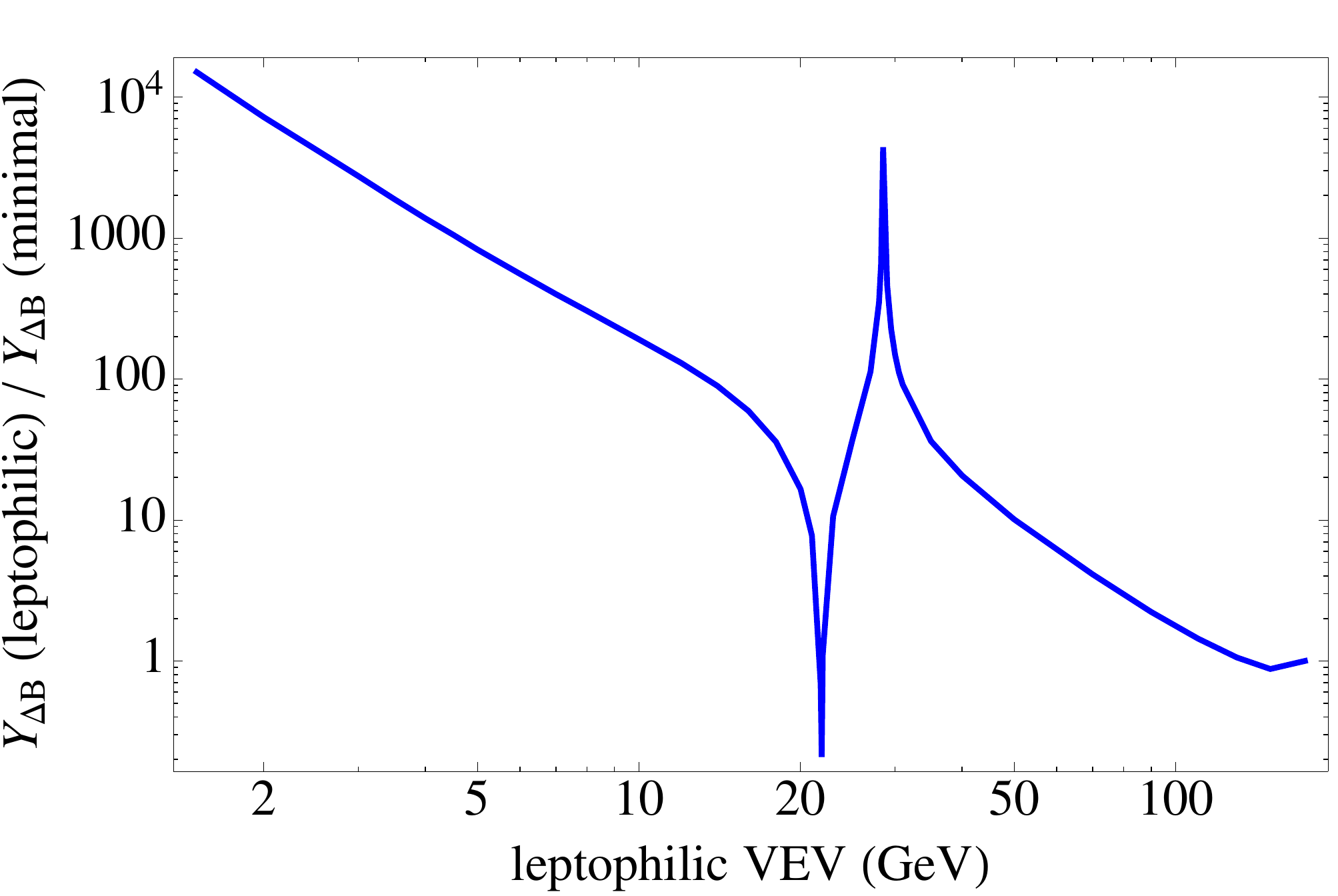}\hspace{0.5cm}\includegraphics[width=0.44\textwidth]{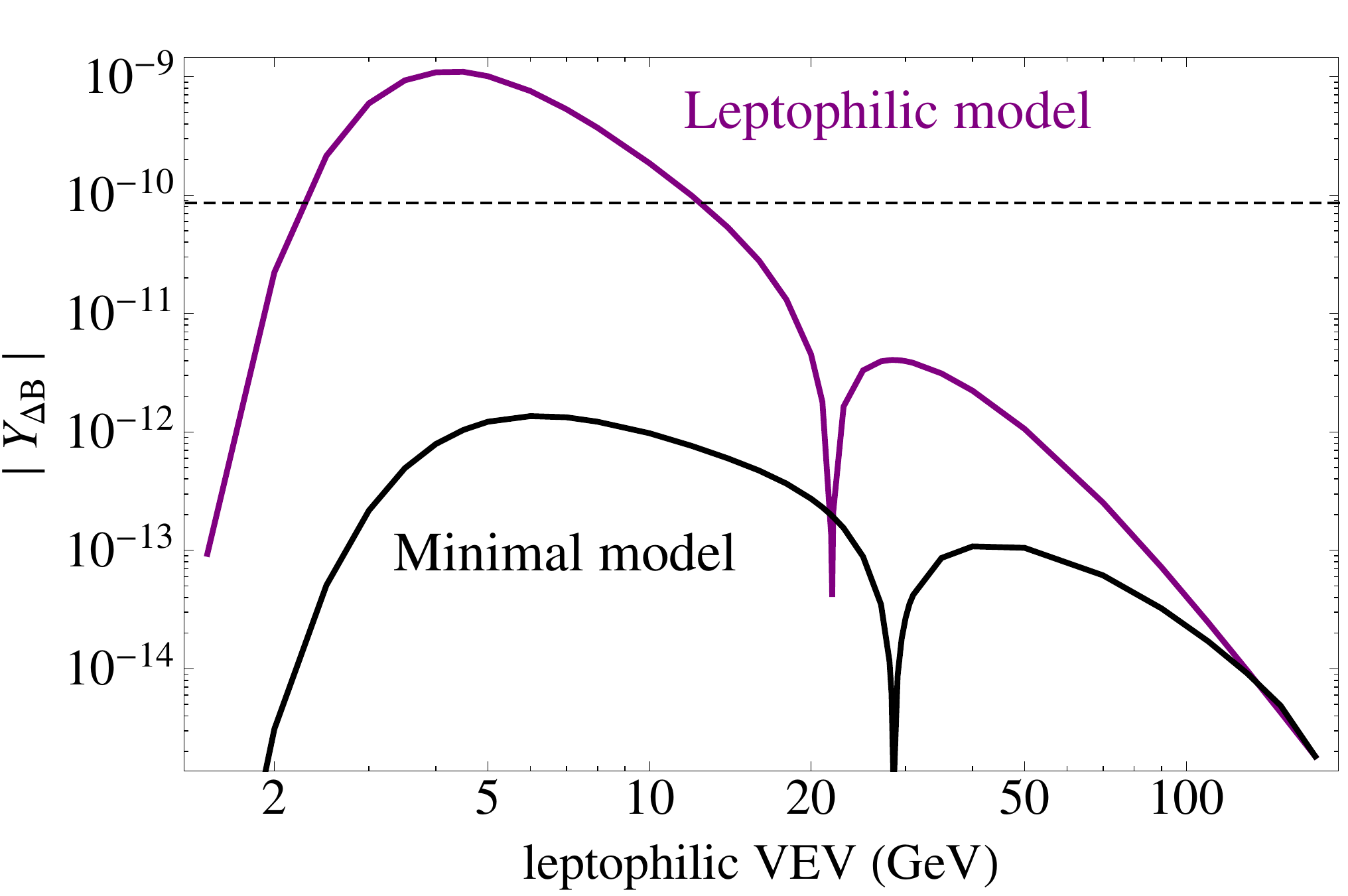}%
\caption{(\textbf{Top}) Ratio of the baryon asymmetry in the leptophilic Higgs model with $\mathrm{Im}\,\omega=1$ to the asymmetry in the minimal, single Higgs model with $\mathrm{Im}\,\omega$ set such that the $L_e$ washout rate is the same in both scenarios for each data point. The enhancement to the asymmetry from changing $\langle\Phi_2\rangle$ is quadratically larger than the tuned, minimal model. (\textbf{Bottom}) Baryon asymmetry in the leptophilic Higgs model (top, purple) and minimal model (bottom, black) with the $L_e$ washout rate the same in both scenarios for each data point. The dashed line indicates the observed baryon asymmetry of the universe. (\textbf{Both}) Other parameters are fixed at $M_N=1$ GeV, $\Delta M_N=1$ GeV, $\eta=\delta=-\mathrm{Re}\,\omega=-\pi/4$.}
\label{fig:vev_ratio}
\end{figure}

To demonstrate the larger asymmetries possible in the leptophilic 2HDM, we compute the baryon asymmetries obtained in both the leptophilic 2HDM and the minimal model, and we show their ratio in the left panel of Fig.~\ref{fig:vev_ratio}. To make a direct comparison, we choose a common set of parameters for both models in Regime III, where the effects are most pronounced: $M_N=1$ GeV, $\Delta M_N=1$ GeV, $\eta=\delta=-\mathrm{Re}\,\omega=-\pi/4$. For the leptophilic 2HDM, we fix $\mathrm{Im}\,\omega=1$ and take as a free parameter the VEV $\langle\Phi_2\rangle$. For the minimal model, we choose the value of $\mathrm{Im}\,\omega$ such that the electron washout rate $\Gamma(L_e\rightarrow N^\dagger)$ is equal to the rate in the leptophilic 2HDM for each $\langle\Phi_2\rangle$. 
The ratio of the baryon asymmetry in the leptophilic 2HDM compared to the minimal model  grows quadratically as $\Phi_2$ decreases from the SM value, as predicted by Eq.~(\ref{eq:generationrate}). 
We see in the right panel of Fig.~\ref{fig:vev_ratio} that enhancements to the baryon asymmetry of $\mathcal{O}(10^3-10^4)$ are possible in the leptophilic Higgs model over the minimal model. The resulting baryon asymmetry in the leptophilic 2HDM is sufficiently large that no mass degeneracy is required to obtain the observed value. Therefore, the leptophilic 2HDM removes the need for any tuning in both the masses of the sterile neutrinos and the Yukawa couplings $F_{\alpha I}$ if $\tan\beta\approx20-80$. The same conclusion holds true for many other values of the $CP$ phases and mixing angles.

%%%%%%%%%%%%%%% FIGURE %%%%%%%%%%%%%%%%%%%%%
%%%%%%%%%%%%%%% FIGURE %%%%%%%%%%%%%%%%%%%%%

One might wonder why such a complete elimination of the tuning in the mass splitting is possible in the leptophilic Higgs model but not in the $\nu$MSM. In other words, why is it now possible to obtain the correct asymmetry with a much shorter coherent oscillation time as compared to the minimal model? The answer is found in the scaling relations, Eqs.~(\ref{eq:generationrate}) and~(\ref{eq:washoutrate}). In the minimal model, one can only increase the Yukawa couplings through alignment, $\cosh(2\mathrm{Im}\,\omega) \gg 1$. But since the asymmetry generation and washout rates scale in the same way with $\cosh(2\mathrm{Im}\,\omega)$, this increase is saturated when the equilibration rate is as early as the electroweak scale. Unfortunately, the baryon asymmetry at the point of saturation is still too small, unless one tunes the sterile-neutrino mass splitting to achieve a longer coherent oscillation time. In our extended model with smaller leptophilic Higgs VEV, the asymmetry generation scales like $\sim\langle\Phi_2\rangle^{-4}$, which rises much faster with a smaller VEV than the washout rate, $\sim\langle\Phi_2\rangle^{-2}$. The point of saturation is still reached in the extended model, but because the asymmetry generation rate increases more than the washout rate, the point of saturation gives a baryon asymmetry which is orders of magnitude larger.

\subsection{Experimental Implications of the Leptophilic Higgs Model}
\label{sec:experiments}

In the leptophilic Higgs $\nu$MSM, which leads to the observed baryon asymmetry without any tuning, there exist new, weakly charged scalars. In a natural theory, the new scalar masses are expected to lie below $\sim$TeV. Thus, experimental studies of the Higgs sector can also act as probes of physics related to leptogenesis.

The baryon asymmetry is largest in models where
\be
\tan\beta \equiv \frac{\langle\Phi_{1}\rangle}{\langle \Phi_2\rangle}\gg1,
\ee
in which the Yukawa couplings of $\Phi_2$ to leptons are enhanced.  In the 2HDM with scalar mixing induced by the potential (\ref{eq:2HDM_potential}), the ratio of VEVs is given by (\ref{eq:tanbeta}), and the $CP$-even mixing angle is
\be
\sin\alpha \approx \frac{\mu_{\rm mix}^2}{m_h^2-\mu_2^2},
\ee
where $m_h^2 = \lambda\langle\Phi_1\rangle^2/4$ and we assume  $\mu_{\rm mix}^2\ll m_h^2 - \mu_{2}^2$. When $\mu_2\approx m_h$, the above approximations break down, and $\sin\alpha \approx 1/\sqrt{2}$.\\

\begin{figure}[t]
\centering
\includegraphics[width=0.4\textwidth]{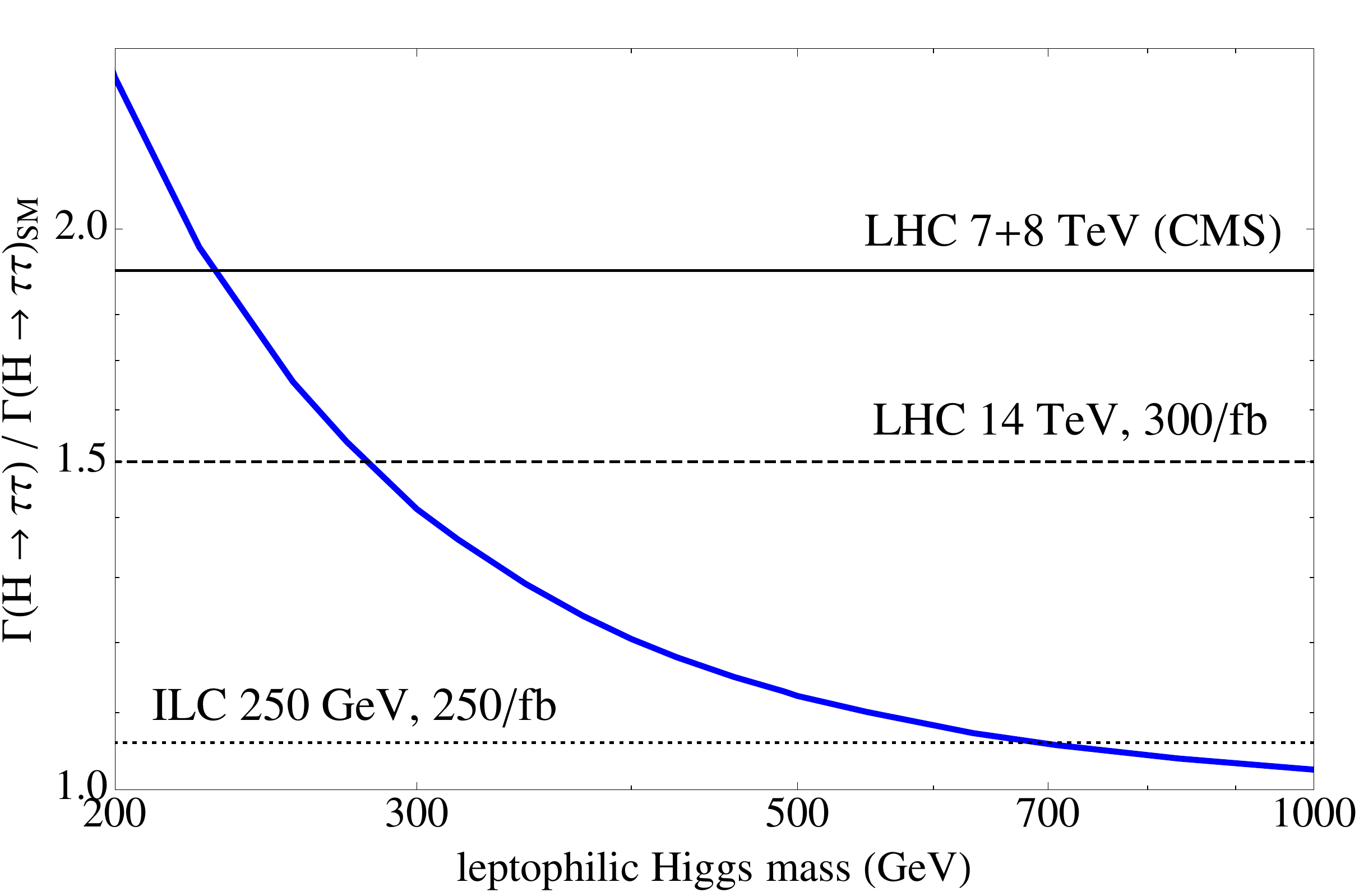}%
\caption{Signal strength of SM Higgs decay to $\tau^+\tau^-$ as a function of the leptophilic Higgs mass $h_\ell$ with $\tan\beta=20$. The enhancement of the $\tau^+\tau^-$ signal strength 
comes from the modification of the SM Higgs coupling to taus (\ref{eq:tauyukawa}). The horizontal solid line is the current CMS 7+8 TeV $2\sigma$ bound \cite{CMS-PAS-HIG-13-004}, and the horizontal dashed (dotted) lines show the $2\sigma$ reach for LHC14 at $300\,\,\mathrm{fb}^{-1}$ (ILC at 250 GeV, $250\,\,\mathrm{fb}^{-1}$). The reach estimates are from \cite{Peskin:2012we}.}
\label{fig:higgs_coupling}
\end{figure}

\noindent {\bf Constraints from SM Higgs searches:} Currently, the strongest constraints on the leptophilic Higgs model are from measurements of the observed SM-like Higgs decays. In the leptophilic model, the SM-like Higgs has a modified  $\tau$ Yukawa coupling
\be\label{eq:tauyukawa}
\lambda_\tau \rightarrow \lambda_\tau\tan\beta\sin\alpha \approx \lambda_\tau \left( \frac{\mu_2^2}{m_h^2-\mu_2^2}\right).
\ee
Even though the SM Higgs doublet does not directly couple to leptons, we see that its coupling to taus is actually enhanced due to a combination of mixing with $\Phi_2$ and the $\tan\beta$ enhancement of the lepton Yukawa couplings. With $\mu_2\sim m_h \approx 126$ GeV, the SM-like Higgs coupling to $\tau^+\tau^-$ is so large that it is excluded by data of SM Higgs decays into taus from the Large Hadron Collider (LHC); the current bound is $\mu_2 \gtrsim220$ GeV. In Fig.~\ref{fig:higgs_coupling}, we show the current exclusion \cite{CMS-PAS-HIG-13-004}, along with the $2\sigma$ reach of the 14 TeV LHC with $300\,\,\mathrm{fb}^{-1}$ of data, and the $2\sigma$ reach of a 250 GeV International Linear Collider (ILC) with $250\,\,\mathrm{fb}^{-1}$ of data \cite{Peskin:2012we}. For this analysis, we calculated the $\Phi_{\rm SM}\rightarrow \tau^+\tau^-$ signal strength with \texttt{2HDMC} \cite{Eriksson:2009ws}.\\

%%%%%%%%%%%%%%% FIGURE %%%%%%%%%%%%%%%%%%%%%
%%%%%%%%%%%%%%% FIGURE %%%%%%%%%%%%%%%%%%%%%

\begin{figure}[t]
\centering
\includegraphics[width=0.4\textwidth]{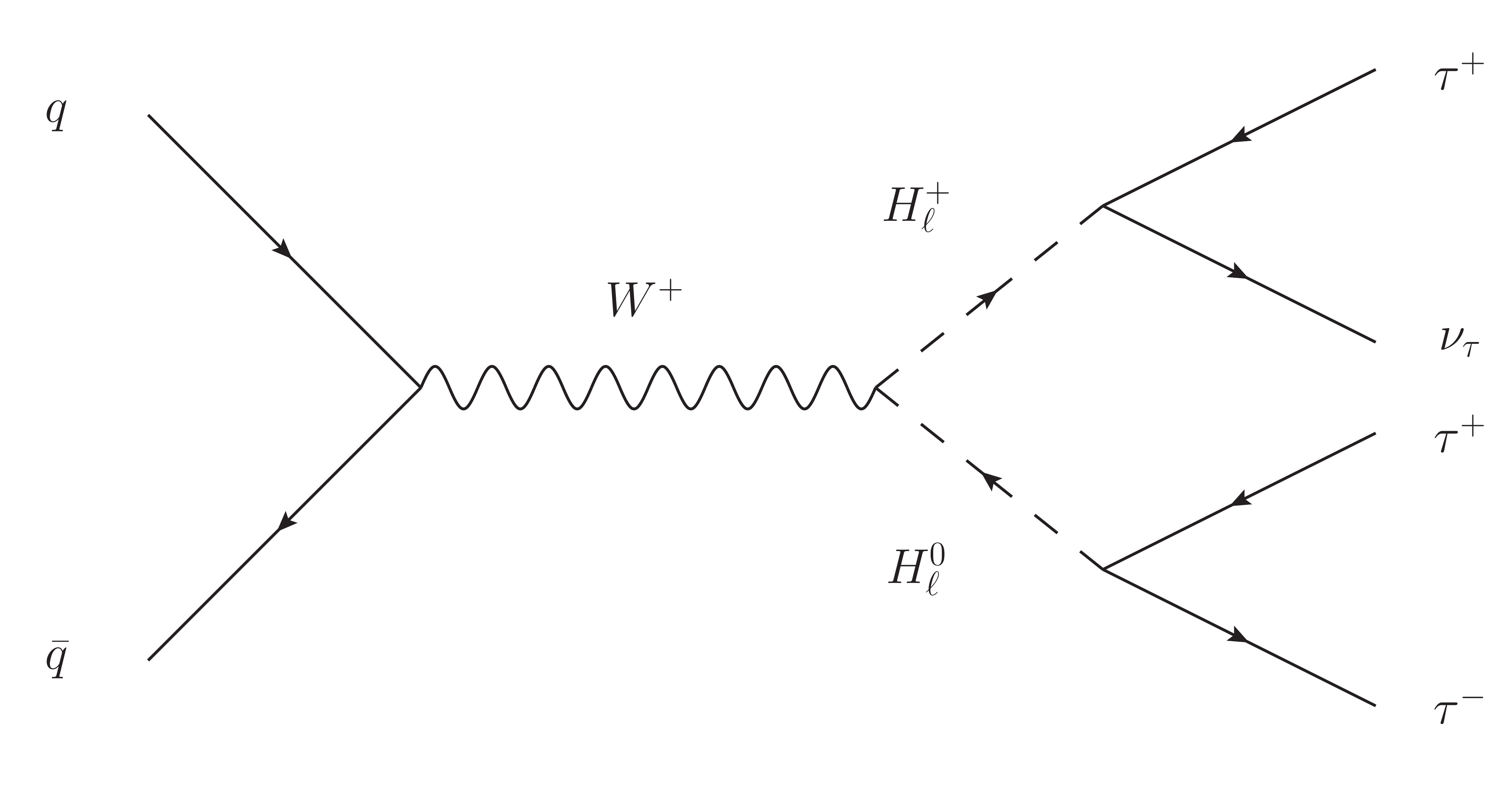}%
\caption{Feynman diagram for production of the leptophilic Higgs states at the LHC and their decays. }
\label{fig:LHC_production}
\end{figure}

\noindent {\bf Direct searches for leptophilic Higgs:} The leptophilic-Higgs-like scalars couple to the electroweak gauge bosons and can be directly produced at colliders. Such searches are currently weaker than the above constraints, but are relevant in extended models where the $\Phi_\ell$ VEV and mixing with the SM are not determined completely by (\ref{eq:2HDM_potential}), and therefore the modification of the SM Higgs coupling is not as large as (\ref{eq:tauyukawa}). Direct searches may also be more relevant for higher luminosities at the LHC. There is one new $CP$-even scalar $H^0_\ell$, a $CP$-odd scalar $A^0_\ell$, and charged scalars $H_\ell^\pm$. The dominant production modes are $pp\rightarrow H_\ell^0/A_\ell^0 + H^\pm_\ell\rightarrow 3\tau+\nu_\tau$ (see Fig.~\ref{fig:LHC_production}).  There is also a $4\tau$ final state, but the production cross section is smaller. The best channel to use in searches for such final states has the same-sign taus decay leptonically and the other tau(s) decay hadronically \cite{Liu:2013gba}; the current constraints from CMS with 8 TeV, 19$\,\,\mathrm{fb}^{-1}$ are $m_{H_\ell}=m_{A_\ell}\lesssim150$ GeV \cite{CMS-PAS-SUS-13-006}. The search in same-sign dileptons + hadronic taus  has a discovery potential of $m_\ell \approx300$ GeV for LHC14 with $\sim200\,\,\mathrm{fb}^{-1}$. Combining this channel with other proposed search modes (such as the all-hadronic channel \cite{Kanemura:2011kx}) could have even higher reach.

%%%%%%%%%%%%%%% FIGURE %%%%%%%%%%%%%%%%%%%%%
%%%%%%%%%%%%%%% FIGURE %%%%%%%%%%%%%%%%%%%%%

Finally, we comment on the possibility that $\Phi_2$  only gives mass   to the neutrinos, while the charged leptons acquire a mass through $\langle\Phi_1\rangle$. In this scenario, the phenomenology changes dramatically; instead of decaying through the large $\tau$ Yukawa interaction, $\Phi_2$ can only decay through either the coupling to sterile neutrinos or the mixing with the SM-like Higgs. The latter is the more likely possibility due to the smallness of the sterile neutrino Yukawas $F_{\alpha I}$, in which case $\Phi_2$ looks exactly like a heavy SM Higgs but with a much smaller single-production cross section and with enhanced pair production. There are no constraints on the leptophilic scalars $H_\ell$ in this scenario, although future searches in the ``golden channel'' $\Phi_{2}\rightarrow 4\ell$ may eventually provide constraints. A linear collider may prove to be a better probe of such final states. If $H_\ell$ instead decays through the Yukawa coupling, then $H_\ell^\pm\rightarrow \ell^\pm +N$, and $H_\ell^\pm$ looks like a slepton decaying to a massless neutralino. The lepton is most likely to be a $\mu$ or $\tau$ because, in the normal hierarchy, these couple strongest to $N$. For the non-tuned models of leptogenesis, the Yukawa couplings are large enough that $H_\ell^\pm$ decays promptly; the slepton bounds constrain $m_{H_\ell}\lesssim300$ GeV  with decays to muons \cite{CMS-PAS-HIG-13-004,ATLAS-CONF-2013-049}, and there are no constraints with decays to taus above the LEP bound of 90 GeV \cite{Abbiendi:2003ji}.

%\newpage
\section{Conclusions}
\label{sec:conclusions}

In this paper, we attempted to provide a comprehensive and coherent overview of the mechanism of baryogenesis through neutrino oscillations. Focusing on the physical time scales involved in the problem rather than the underlying model parameters, we identified three broad regimes depending on the relative ordering of this time scales (the neutrino oscillation time scale, the equilibration time scale, and the sphaleron decoupling time scale). While these regimes are not new - they have been identified in past works~\cite{Asaka:2005pn,Shaposhnikov:2008pf,Asaka:2010kk,Canetti:2010aw,Asaka:2011wq,Drewes:2012ma,Canetti:2012kh} either through scans or with individual points - our work endeavours to clarify the physical basis for these regimes and for their interconnectedness. On a more quantitative level, our calculation also includes an improvement upon previous calculations of the baryon asymmetry by including the effects of scatterings between left-handed leptons and the thermal bath during asymmetry generation.

One of the less appealing features of this mechanism is the need to fine-tune some of the model parameters. We showed that this is fundamentally related to a certain coincidence required of the physical time scales and that a tuning of no less than one part in $10^5$ is necessary throughout the parameter space. In one regime the tuning is entirely in the mass terms of the sterile neutrinos, while in other regimes it is also manifested strongly in a certain alignment of the Yukawa couplings, which allows them to be much larger than what the see-saw relation, Eq.~(\ref{eq:seesaw_lag}), would na\"ively imply. 

The fine-tuning of model parameters is an unavoidable feature of the minimal model. In the last part of this work we considered an extended model with an additional electroweak Higgs boson that predominantly couples to the sterile neutrinos and SM leptons. We showed that the correct baryon asymmetry can be obtained with no tuning of the sterile neutrino parameters if the leptophilic Higgs boson has a VEV of order $\sim$ GeV. This extra Higgs boson can be searched for, and discovered, at the LHC.

\begin{acknowledgments}

We would like to thank Eder Izaguirre, Gordan Krnjaic, Maxim Pospelov, and Carlos Tamarit for helpful discussions. BS is supported in part by the Canadian Institute of Particle Physics.  IY is supported in part by funds from the Natural Sciences and Engineering Research Council (NSERC) of Canada. This research was supported in part by Perimeter Institute for Theoretical Physics. Research at Perimeter Institute is supported by the Government of Canada through Industry Canada and by the Province of Ontario through the Ministry of Research and Innovation.
\end{acknowledgments}

%\newpage
%\onecolumngrid
\appendix

%%%%%%%%%%%%%%%%
% Appendix A
%%%%%%%%%%%%%%%%

\renewcommand{\theequation}{A-\arabic{equation}}
\setcounter{equation}{0}

\section{Casas-Ibarra parameterization}
\label{sec:casasibarra}
The Yukawa coupling $F_{\alpha I}$ in Eq.~(\ref{eq:typei}) can be written as
\be
F_{\alpha I} = \frac{i}{\langle\Phi\rangle} U_{\nu}\sqrt{m_\nu} R \sqrt{M_N},
\ee
where $m_\nu$ is the diagonal matrix of LH neutrino masses as determined from oscillation data, $U_\nu$ is the MNS LH neutrino mixing matrix, and $M_N$ is the diagonal matrix of sterile neutrino masses. For the normal hierarchy, and with $m_1=0$, the decomposition of $U_\nu$ and $R$ are
\bea
U_\nu &=& \left(\begin{array}{ccc} 
1 & 0 & 0 \\
0 & \cos\theta_{23} & \sin\theta_{23} \\
0 & -\sin\theta_{23} & \cos\theta_{23} \end{array}\right)
\mathrm{diag}\left( e^{i\delta/2},\,1,\,e^{-i\delta/2}\right)\nonumber\\
&& {}\times\left(\begin{array}{ccc} 
\cos\theta_{13} & 0 & \sin\theta_{13} \\
0 & 1 & 0 \\
-\sin\theta_{13} & 0 & \cos\theta_{13} \end{array}\right)
\mathrm{diag}\left( e^{-i\delta/2},\,1,\,e^{i\delta/2}\right)\nonumber\\
&& {}\times
\left(\begin{array}{ccc} 
\cos\theta_{12} & \sin\theta_{12} & 0 \\
-\sin\theta_{12} & \cos\theta_{12} & 0 \\
0 & 0 & 1 \end{array}\right)
\mathrm{diag}\left( 1,\,e^{-i\eta},\,1\right),\\
R&=& \left(\begin{array}{cc} 
0 & 0 \\
\cos\omega & \sin\omega \\
-\sin\omega & \cos\omega
\end{array}\right).
\eea
The $\theta_{ij}$ are the usual mixing angles. In general, there is another Majorana phase appearing in $U_\nu$, but it only appears in terms proportional to $m_1$, which we assume to be zero as explained in Section~\ref{sec:models}.

%%%%%%%%%%%%%%%%%%%%%%%%%%%%%%%%%%%%%%%%%%%%%%%%%%%%%%%%%%%%%%%%%%%%%%%%%%%%
%%%%%%%%%%%%%%%%%%%%%%%%%%%%%%%%%%%%%%%%%%%%%%%%%%%%%%%%%%%%%%%%%%%%%%%%%%%%
%%%%%%%%%%%%%%%%%%%%%%%%%%%%%%%%%%%%%%%%%%%%%%%%%%%%%%%%%%%%%%%%%%%%%%%%%%%%

%%%%%%%%%%%%%%%%
% Appendix B
%%%%%%%%%%%%%%%%

\renewcommand{\theequation}{B-\arabic{equation}}
\setcounter{equation}{0}

\section{Full density matrix evolution equations}
\label{sec:fullevolution}
In Section \ref{sec:calculation}, we outlined the basic method for calculating the baryon asymmetry, providing schematic evolution equations for $\rho_N$, $\rho_{\bar N}$, and $\rho_{L-\bar{L}}$ in Eqs.~(\ref{eq:evol_first})-(\ref{eq:evol_lepton}). It turns out to be simpler to move to a different basis for the sterile neutrino density matrices \cite{Shaposhnikov:2008pf}:
\bea
\delta\rho_+ &=& \frac{\rho_N + \rho_{\bar N}}{2} - \rho_L^{\rm eq}\,\mathbb{I}_{2\times2},\\
\delta\rho_- &=& \rho_N-\rho_{\bar N}.
\eea
In this basis, the sterile neutrinos are expressed in terms of \emph{deviations from equilibrium}: $\delta\rho_+$ is the sterile $CP$-even deviation from equilibrium, while $\delta\rho_-$ is the sterile $CP$-odd deviation from equilibrium. This basis has the nice property that, in thermal equilibrium, $\delta\rho_+$, $\delta\rho_-$, and $\rho_{B-3L_\alpha}$ all vanish. Furthermore, when looking at individual terms that generate an asymmetry, it can be convenient to separate out the action of $CP$ violation on sterile neutrino oscillations themselves, which source $\delta\rho_-$, and the presence of $CP$ violation in $L-N$ scattering, which involves  $\mathrm{Im}\,\omega$, as well as the Dirac and Majorana $CP$ phases from the MNS matrix.

%%%%%%%%%%%%%%%% FIGURE %%%%%%%%%%%%%%%%%%%%%
\begin{figure}[t]
\centering
\includegraphics[width=0.4\textwidth]{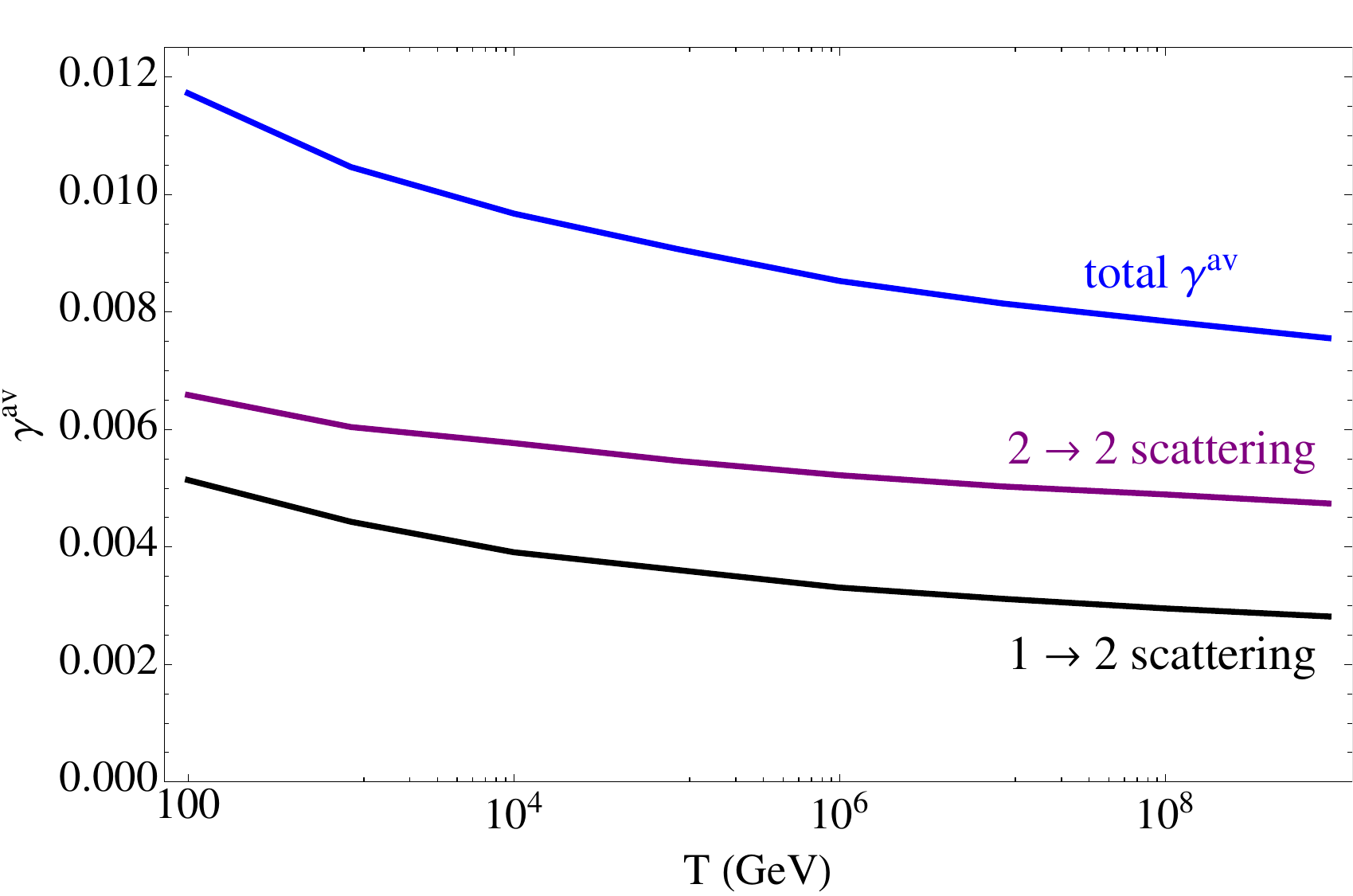}%
\caption{Temperature dependence of the coefficient $\gamma^{\rm av}(T)$ in the sterile neutrino production rate as defined in Eq.~(\ref{eq:gamma_avg_def}). We show the contributions from $1\leftrightarrow2$ scattering, $2\rightarrow2$ scattering, and the total leading order rate. All values are taken from \cite{Besak:2012qm}.}
\label{fig:gamma_avg}
\end{figure}
%%%%%%%%%%%%%%%% FIGURE %%%%%%%%%%%%%%%%%%%%%

Before providing the full kinetic equations used in the paper, we highlight an additional modification of the evolution equations from \cite{Asaka:2011wq}. In section~\ref{sec:calculation} we parametrized the rate $\Gamma(L\rightarrow N^\dagger)$ in Eq.~(\ref{eq:gamma_avg_def}) with the aid of the function $\gamma_{\rm av}(T)$. This function is shown in Fig.~(\ref{fig:gamma_avg}). Asaka and Ishida observed that the assumption that the $N$ production and $L$ destruction rates are parameterized by the same $\gamma^{\rm av}(T)$ is erroneous: while at leading order, $N$ production always proceeds from SM initial states, some $L$ destruction processes have sterile neutrinos in the initial state and are consequently suppressed by the out-of-equilibrium $N$ phase space density. For both $2\rightarrow2$ and $1\leftrightarrow2$ scattering processes, 1/3 of the lepton destruction processes are suppressed by a factor of $\rho_N$.

The various contributions to $\gamma^{\rm av}(T)$ can be extracted  from the analysis in \cite{Besak:2012qm}. We then modify the lepton destruction rate such that the coefficient  changes as
\be
 \gamma^{\rm av}(T)\rightarrow \frac{2}{3}\gamma^{\rm av}(T) + \frac{\rho_N+\rho_{\bar N}}{6}\gamma^{\rm av}(T).
 \ee
Putting together all of the modifications due to the rotation to the $\delta\rho_+$, $\delta\rho_-$ basis, the transformation to the $B-3L_\alpha$ lepton charges, and the above modifications to the scattering rates, we have
\onecolumngrid
\small
\bea
i\frac{d\delta\rho_+}{dt} &=& [ \mathrm{Re} H_N,\, \delta\rho_+]-\frac{i}{2}\{\mathrm{Re}\Gamma_N,\,\delta\rho_+\} - \frac{iT}{4}\gamma^{\rm av}(T)(F^\dagger\rho_{L-\bar L} F-F^{\rm T}\rho_{L-\bar L} F^*)\nonumber\\
&&{}-\frac{iT}{24}\gamma^{\rm av}(T)\left[\{F^\dagger\rho_{L-\bar L}F,\,2\delta\rho_++\delta\rho_-\}-\{F^{\rm T}\rho_{L-\bar L}F^*,\,2\delta\rho_+-\delta\rho_-\}\right]\nonumber\\
&&{} + \frac{i}{2}[\mathrm{Im}H_N,\,\delta\rho_-] + \frac{1}{4}\{\mathrm{Im}\Gamma_N,\,\delta\rho_-\},\\
i\frac{d\delta\rho_-}{dt} &=& [ \mathrm{Re} H_N,\, \delta\rho_-]-\frac{i}{2}\{\mathrm{Re}\Gamma_N,\,\delta\rho_-\} - \frac{iT}{2}\gamma^{\rm av}(T)(F^\dagger\rho_{L-\bar L} F+F^{\rm T}\rho_{L-\bar L} F^*)\nonumber\\
&&{}-\frac{iT}{12}\gamma^{\rm av}(T)\left[\{F^\dagger\rho_{L-\bar L}F,\,2\delta\rho_++\delta\rho_-\}+\{F^{\rm T}\rho_{L-\bar L}F^*,\,2\delta\rho_+-\delta\rho_-\}\right]\nonumber\\
&&{} + 2i[\mathrm{Im}H_N,\,\delta\rho_+] + \{\mathrm{Im}\Gamma_N,\,\delta\rho_+\},\\
i\frac{d\rho_{B-3L}}{dt}&=&\frac{3i}{2}\{\Gamma_L,\,\rho_{L-\bar L}\}+\frac{iT}{2}\gamma^{\rm av}(T)\left[F(2\delta\rho_++\delta\rho_-)F^\dagger+F^*(2\delta\rho_+-\delta\rho_-)F^{\rm T}\right]\rho_{L-\bar L}\nonumber\\
&& {} - \frac{3iT}{2}\gamma^{\rm av}(T)\left[F(2\delta\rho_++\delta\rho_-)F^\dagger-F^*(2\delta\rho_+-\delta\rho_-)F^{\rm T}\right].
\eea
\normalsize
\twocolumngrid
Here, we have defined
\bea
(\Gamma_N)_{IJ} &=& \gamma^{\rm av}\,T (F^\dagger F)_{IJ},\\
(H_N)_{IJ} &=& \frac{T}{8}(F^\dagger F)_{IJ} + \frac{M_I^2-M_J^2}{2T},\\
(\Gamma_L)_{\alpha\beta} &=& \gamma^{\rm av}\,T\,\mathrm{Re}(FF^\dagger)_{\alpha\beta},
\eea
and $\rho_{L-\bar L}$ can be expressed in terms of $\rho_{B-3L}$ according to Eq.~(\ref{eq:chem_relations}).

As noted in \cite{Canetti:2010aw}, this system of differential equations is ``stiff'' due to the fact that the sterile neutrino oscillation frequency increases monotonically with time, and integrating from an early time to $t_{\rm W}$ can be very computationally intensive. Fortunately, there is a workaround: washout cannot equilibrate much before the weak scale or else the asymmetry would be entirely wiped out, while oscillation frequencies are very fast at $t_{\rm W}$ only when the oscillations began at a much earlier time. This separation of $t_{\rm osc}\ll t_{\rm W}$ allow us to divide the equation evolution into two regimes. At early times, when the oscillations are responsible for generating the lepton asymmetries, we solve the full kinetic equations above. At late times, the oscillations are rapid and the contributions to the asymmetry average to zero, and so we solve the same system of equations \emph{except} that we set the off-diagonal components to zero. The solutions are matched at a time $t_{\rm match}=500t_{\rm osc}$, where
\be
t_{\rm osc} \approx \frac{(M_{\rm Pl}/1.66\sqrt{g_*})^{1/3}}{2(M_3^2-M_2^2)^{2/3}}
\ee
is the time when oscillations become rapid. We have checked numerically that at least 90\% of the asymmetry is generated by $t_{\rm match}$, and we have also done extensive numerical checks of the validity of this two-stage solution. When $t_{\rm osc}\sim t_{\rm W}$, then it is fast to integrate the full equations right up to $t_{\rm W}$.

Integration can also be slow when we include an interpolating function representation of the numerical coefficients $\gamma^{\rm av}(T)$ for the solution of the full equations. Because these functions vary slowly with time, and the rates are most relevant at the times of asymmetry generation and washout, we set $\gamma^{\rm av} = \gamma^{\rm av}(t_{\rm osc})$ for the first stage of the solution. In the second stage of the solution, we incorporate the full time dependence of $\gamma^{\rm av}(T)$.

\bibliography{Lepto_nuoscillation-bib}

\begin{thebibliography}{35}%
\makeatletter
\providecommand \@ifxundefined [1]{%
 \@ifx{#1\undefined}
}%
\providecommand \@ifnum [1]{%
 \ifnum #1\expandafter \@firstoftwo
 \else \expandafter \@secondoftwo
 \fi
}%
\providecommand \@ifx [1]{%
 \ifx #1\expandafter \@firstoftwo
 \else \expandafter \@secondoftwo
 \fi
}%
\providecommand \natexlab [1]{#1}%
\providecommand \enquote  [1]{``#1''}%
\providecommand \bibnamefont  [1]{#1}%
\providecommand \bibfnamefont [1]{#1}%
\providecommand \citenamefont [1]{#1}%
\providecommand \href@noop [0]{\@secondoftwo}%
\providecommand \href [0]{\begingroup \@sanitize@url \@href}%
\providecommand \@href[1]{\@@startlink{#1}\@@href}%
\providecommand \@@href[1]{\endgroup#1\@@endlink}%
\providecommand \@sanitize@url [0]{\catcode `\\12\catcode `\$12\catcode
  `\&12\catcode `\#12\catcode `\^12\catcode `\_12\catcode `\%12\relax}%
\providecommand \@@startlink[1]{}%
\providecommand \@@endlink[0]{}%
\providecommand \url  [0]{\begingroup\@sanitize@url \@url }%
\providecommand \@url [1]{\endgroup\@href {#1}{\urlprefix }}%
\providecommand \urlprefix  [0]{URL }%
\providecommand \Eprint [0]{\href }%
\providecommand \doibase [0]{http://dx.doi.org/}%
\providecommand \selectlanguage [0]{\@gobble}%
\providecommand \bibinfo  [0]{\@secondoftwo}%
\providecommand \bibfield  [0]{\@secondoftwo}%
\providecommand \translation [1]{[#1]}%
\providecommand \BibitemOpen [0]{}%
\providecommand \bibitemStop [0]{}%
\providecommand \bibitemNoStop [0]{.\EOS\space}%
\providecommand \EOS [0]{\spacefactor3000\relax}%
\providecommand \BibitemShut  [1]{\csname bibitem#1\endcsname}%
\let\auto@bib@innerbib\@empty
%</preamble>
\bibitem [{\citenamefont {Akhmedov}\ \emph {et~al.}(1998)\citenamefont
  {Akhmedov}, \citenamefont {Rubakov},\ and\ \citenamefont
  {Smirnov}}]{Akhmedov:1998qx}%
  \BibitemOpen
  \bibfield  {author} {\bibinfo {author} {\bibfnamefont {E.~K.}\ \bibnamefont
  {Akhmedov}}, \bibinfo {author} {\bibfnamefont {V.}~\bibnamefont {Rubakov}}, \
  and\ \bibinfo {author} {\bibfnamefont {A.~Y.}\ \bibnamefont {Smirnov}},\
  }\href {\doibase 10.1103/PhysRevLett.81.1359} {\bibfield  {journal} {\bibinfo
   {journal} {Phys.Rev.Lett.}\ }\textbf {\bibinfo {volume} {81}},\ \bibinfo
  {pages} {1359} (\bibinfo {year} {1998})},\ \Eprint
  {http://arxiv.org/abs/hep-ph/9803255} {arXiv:hep-ph/9803255 [hep-ph]}
  \BibitemShut {NoStop}%
%%CITATION = HEP-PH/9803255;%%
\bibitem [{\citenamefont {Asaka}\ and\ \citenamefont
  {Shaposhnikov}(2005)}]{Asaka:2005pn}%
  \BibitemOpen
  \bibfield  {author} {\bibinfo {author} {\bibfnamefont {T.}~\bibnamefont
  {Asaka}}\ and\ \bibinfo {author} {\bibfnamefont {M.}~\bibnamefont
  {Shaposhnikov}},\ }\href {\doibase 10.1016/j.physletb.2005.06.020} {\bibfield
   {journal} {\bibinfo  {journal} {Phys.Lett.}\ }\textbf {\bibinfo {volume}
  {B620}},\ \bibinfo {pages} {17} (\bibinfo {year} {2005})},\ \Eprint
  {http://arxiv.org/abs/hep-ph/0505013} {arXiv:hep-ph/0505013 [hep-ph]}
  \BibitemShut {NoStop}%
%%CITATION = HEP-PH/0505013;%%
\bibitem [{\citenamefont {Asaka}\ \emph {et~al.}(2005)\citenamefont {Asaka},
  \citenamefont {Blanchet},\ and\ \citenamefont {Shaposhnikov}}]{Asaka:2005an}%
  \BibitemOpen
  \bibfield  {author} {\bibinfo {author} {\bibfnamefont {T.}~\bibnamefont
  {Asaka}}, \bibinfo {author} {\bibfnamefont {S.}~\bibnamefont {Blanchet}}, \
  and\ \bibinfo {author} {\bibfnamefont {M.}~\bibnamefont {Shaposhnikov}},\
  }\href {\doibase 10.1016/j.physletb.2005.09.070} {\bibfield  {journal}
  {\bibinfo  {journal} {Phys.Lett.}\ }\textbf {\bibinfo {volume} {B631}},\
  \bibinfo {pages} {151} (\bibinfo {year} {2005})},\ \Eprint
  {http://arxiv.org/abs/hep-ph/0503065} {arXiv:hep-ph/0503065 [hep-ph]}
  \BibitemShut {NoStop}%
%%CITATION = HEP-PH/0503065;%%
\bibitem [{\citenamefont {Laine}\ and\ \citenamefont
  {Shaposhnikov}(2008)}]{Laine:2008pg}%
  \BibitemOpen
  \bibfield  {author} {\bibinfo {author} {\bibfnamefont {M.}~\bibnamefont
  {Laine}}\ and\ \bibinfo {author} {\bibfnamefont {M.}~\bibnamefont
  {Shaposhnikov}},\ }\href {\doibase 10.1088/1475-7516/2008/06/031} {\bibfield
  {journal} {\bibinfo  {journal} {JCAP}\ }\textbf {\bibinfo {volume} {0806}},\
  \bibinfo {pages} {031} (\bibinfo {year} {2008})},\ \Eprint
  {http://arxiv.org/abs/0804.4543} {arXiv:0804.4543 [hep-ph]} \BibitemShut
  {NoStop}%
%%CITATION = ARXIV:0804.4543;%%
\bibitem [{\citenamefont {Canetti}\ \emph {et~al.}(2013)\citenamefont
  {Canetti}, \citenamefont {Drewes}, \citenamefont {Frossard},\ and\
  \citenamefont {Shaposhnikov}}]{Canetti:2012kh}%
  \BibitemOpen
  \bibfield  {author} {\bibinfo {author} {\bibfnamefont {L.}~\bibnamefont
  {Canetti}}, \bibinfo {author} {\bibfnamefont {M.}~\bibnamefont {Drewes}},
  \bibinfo {author} {\bibfnamefont {T.}~\bibnamefont {Frossard}}, \ and\
  \bibinfo {author} {\bibfnamefont {M.}~\bibnamefont {Shaposhnikov}},\ }\href
  {\doibase 10.1103/PhysRevD.87.093006} {\bibfield  {journal} {\bibinfo
  {journal} {Phys.Rev.}\ }\textbf {\bibinfo {volume} {D87}},\ \bibinfo {pages}
  {093006} (\bibinfo {year} {2013})},\ \Eprint {http://arxiv.org/abs/1208.4607}
  {arXiv:1208.4607 [hep-ph]} \BibitemShut {NoStop}%
%%CITATION = ARXIV:1208.4607;%%
\bibitem [{\citenamefont {Shaposhnikov}(2008)}]{Shaposhnikov:2008pf}%
  \BibitemOpen
  \bibfield  {author} {\bibinfo {author} {\bibfnamefont {M.}~\bibnamefont
  {Shaposhnikov}},\ }\href {\doibase 10.1088/1126-6708/2008/08/008} {\bibfield
  {journal} {\bibinfo  {journal} {JHEP}\ }\textbf {\bibinfo {volume} {0808}},\
  \bibinfo {pages} {008} (\bibinfo {year} {2008})},\ \Eprint
  {http://arxiv.org/abs/0804.4542} {arXiv:0804.4542 [hep-ph]} \BibitemShut
  {NoStop}%
%%CITATION = ARXIV:0804.4542;%%
\bibitem [{\citenamefont {Asaka}\ and\ \citenamefont
  {Ishida}(2010)}]{Asaka:2010kk}%
  \BibitemOpen
  \bibfield  {author} {\bibinfo {author} {\bibfnamefont {T.}~\bibnamefont
  {Asaka}}\ and\ \bibinfo {author} {\bibfnamefont {H.}~\bibnamefont {Ishida}},\
  }\href {\doibase 10.1016/j.physletb.2010.07.016} {\bibfield  {journal}
  {\bibinfo  {journal} {Phys.Lett.}\ }\textbf {\bibinfo {volume} {B692}},\
  \bibinfo {pages} {105} (\bibinfo {year} {2010})},\ \Eprint
  {http://arxiv.org/abs/1004.5491} {arXiv:1004.5491 [hep-ph]} \BibitemShut
  {NoStop}%
%%CITATION = ARXIV:1004.5491;%%
\bibitem [{\citenamefont {Canetti}\ and\ \citenamefont
  {Shaposhnikov}(2010)}]{Canetti:2010aw}%
  \BibitemOpen
  \bibfield  {author} {\bibinfo {author} {\bibfnamefont {L.}~\bibnamefont
  {Canetti}}\ and\ \bibinfo {author} {\bibfnamefont {M.}~\bibnamefont
  {Shaposhnikov}},\ }\href {\doibase 10.1088/1475-7516/2010/09/001} {\bibfield
  {journal} {\bibinfo  {journal} {JCAP}\ }\textbf {\bibinfo {volume} {1009}},\
  \bibinfo {pages} {001} (\bibinfo {year} {2010})},\ \Eprint
  {http://arxiv.org/abs/1006.0133} {arXiv:1006.0133 [hep-ph]} \BibitemShut
  {NoStop}%
%%CITATION = ARXIV:1006.0133;%%
\bibitem [{\citenamefont {Asaka}\ \emph {et~al.}(2012)\citenamefont {Asaka},
  \citenamefont {Eijima},\ and\ \citenamefont {Ishida}}]{Asaka:2011wq}%
  \BibitemOpen
  \bibfield  {author} {\bibinfo {author} {\bibfnamefont {T.}~\bibnamefont
  {Asaka}}, \bibinfo {author} {\bibfnamefont {S.}~\bibnamefont {Eijima}}, \
  and\ \bibinfo {author} {\bibfnamefont {H.}~\bibnamefont {Ishida}},\ }\href
  {\doibase 10.1088/1475-7516/2012/02/021} {\bibfield  {journal} {\bibinfo
  {journal} {JCAP}\ }\textbf {\bibinfo {volume} {1202}},\ \bibinfo {pages}
  {021} (\bibinfo {year} {2012})},\ \Eprint {http://arxiv.org/abs/1112.5565}
  {arXiv:1112.5565 [hep-ph]} \BibitemShut {NoStop}%
%%CITATION = ARXIV:1112.5565;%%
\bibitem [{\citenamefont {Drewes}\ and\ \citenamefont
  {Garbrecht}(2013)}]{Drewes:2012ma}%
  \BibitemOpen
  \bibfield  {author} {\bibinfo {author} {\bibfnamefont {M.}~\bibnamefont
  {Drewes}}\ and\ \bibinfo {author} {\bibfnamefont {B.}~\bibnamefont
  {Garbrecht}},\ }\href {\doibase 10.1007/JHEP03(2013)096} {\bibfield
  {journal} {\bibinfo  {journal} {JHEP}\ }\textbf {\bibinfo {volume} {1303}},\
  \bibinfo {pages} {096} (\bibinfo {year} {2013})},\ \Eprint
  {http://arxiv.org/abs/1206.5537} {arXiv:1206.5537 [hep-ph]} \BibitemShut
  {NoStop}%
%%CITATION = ARXIV:1206.5537;%%
\bibitem [{\citenamefont {Drewes}(2013)}]{Drewes:2013gca}%
  \BibitemOpen
  \bibfield  {author} {\bibinfo {author} {\bibfnamefont {M.}~\bibnamefont
  {Drewes}},\ }\href {\doibase 10.1142/S0218301313300191} {\bibfield  {journal}
  {\bibinfo  {journal} {Int.J.Mod.Phys.}\ }\textbf {\bibinfo {volume} {E22}},\
  \bibinfo {pages} {1330019} (\bibinfo {year} {2013})},\ \Eprint
  {http://arxiv.org/abs/1303.6912} {arXiv:1303.6912 [hep-ph]} \BibitemShut
  {NoStop}%
%%CITATION = ARXIV:1303.6912;%%
\bibitem [{\citenamefont {Minkowski}(1977)}]{Minkowski:1977sc}%
  \BibitemOpen
  \bibfield  {author} {\bibinfo {author} {\bibfnamefont {P.}~\bibnamefont
  {Minkowski}},\ }\href {\doibase 10.1016/0370-2693(77)90435-X} {\bibfield
  {journal} {\bibinfo  {journal} {Phys.Lett.}\ }\textbf {\bibinfo {volume}
  {B67}},\ \bibinfo {pages} {421} (\bibinfo {year} {1977})}\BibitemShut
  {NoStop}%
%%CITATION = PHLTA,B67,421;%%
\bibitem [{\citenamefont {Kuzmin}\ \emph {et~al.}(1985)\citenamefont {Kuzmin},
  \citenamefont {Rubakov},\ and\ \citenamefont {Shaposhnikov}}]{Kuzmin:1985mm}%
  \BibitemOpen
  \bibfield  {author} {\bibinfo {author} {\bibfnamefont {V.}~\bibnamefont
  {Kuzmin}}, \bibinfo {author} {\bibfnamefont {V.}~\bibnamefont {Rubakov}}, \
  and\ \bibinfo {author} {\bibfnamefont {M.}~\bibnamefont {Shaposhnikov}},\
  }\href {\doibase 10.1016/0370-2693(85)91028-7} {\bibfield  {journal}
  {\bibinfo  {journal} {Phys.Lett.}\ }\textbf {\bibinfo {volume} {B155}},\
  \bibinfo {pages} {36} (\bibinfo {year} {1985})}\BibitemShut {NoStop}%
%%CITATION = PHLTA,B155,36;%%
\bibitem [{\citenamefont {D'Onofrio}\ \emph {et~al.}(2012)\citenamefont
  {D'Onofrio}, \citenamefont {Rummukainen},\ and\ \citenamefont
  {Tranberg}}]{D'Onofrio:2012jk}%
  \BibitemOpen
  \bibfield  {author} {\bibinfo {author} {\bibfnamefont {M.}~\bibnamefont
  {D'Onofrio}}, \bibinfo {author} {\bibfnamefont {K.}~\bibnamefont
  {Rummukainen}}, \ and\ \bibinfo {author} {\bibfnamefont {A.}~\bibnamefont
  {Tranberg}},\ }\href {\doibase 10.1007/JHEP08(2012)123} {\bibfield  {journal}
  {\bibinfo  {journal} {JHEP}\ }\textbf {\bibinfo {volume} {1208}},\ \bibinfo
  {pages} {123} (\bibinfo {year} {2012})},\ \Eprint
  {http://arxiv.org/abs/1207.0685} {arXiv:1207.0685 [hep-ph]} \BibitemShut
  {NoStop}%
%%CITATION = ARXIV:1207.0685;%%
\bibitem [{\citenamefont {Sakharov}(1967)}]{Sakharov:1967dj}%
  \BibitemOpen
  \bibfield  {author} {\bibinfo {author} {\bibfnamefont {A.}~\bibnamefont
  {Sakharov}},\ }\href {\doibase 10.1070/PU1991v034n05ABEH002497} {\bibfield
  {journal} {\bibinfo  {journal} {Pisma Zh.Eksp.Teor.Fiz.}\ }\textbf {\bibinfo
  {volume} {5}},\ \bibinfo {pages} {32} (\bibinfo {year} {1967})}\BibitemShut
  {NoStop}%
%%CITATION = ZFPRA,5,32;%%
\bibitem [{\citenamefont {Kolb}\ and\ \citenamefont
  {Turner}(1990)}]{Kolb:1990vq}%
  \BibitemOpen
  \bibfield  {author} {\bibinfo {author} {\bibfnamefont {E.~W.}\ \bibnamefont
  {Kolb}}\ and\ \bibinfo {author} {\bibfnamefont {M.~S.}\ \bibnamefont
  {Turner}},\ }\href@noop {} {\bibfield  {journal} {\bibinfo  {journal}
  {Front.Phys.}\ }\textbf {\bibinfo {volume} {69}},\ \bibinfo {pages} {1}
  (\bibinfo {year} {1990})}\BibitemShut {NoStop}%
%%CITATION = FRPHA,69,1;%%
\bibitem [{\citenamefont {Beringer}\ \emph {et~al.}(2012)\citenamefont
  {Beringer} \emph {et~al.}}]{Beringer:1900zz}%
  \BibitemOpen
  \bibfield  {author} {\bibinfo {author} {\bibfnamefont {J.}~\bibnamefont
  {Beringer}} \emph {et~al.} (\bibinfo {collaboration} {Particle Data Group}),\
  }\href {\doibase 10.1103/PhysRevD.86.010001} {\bibfield  {journal} {\bibinfo
  {journal} {Phys.Rev.}\ }\textbf {\bibinfo {volume} {D86}},\ \bibinfo {pages}
  {010001} (\bibinfo {year} {2012})}\BibitemShut {NoStop}%
%%CITATION = PHRVA,D86,010001;%%
\bibitem [{\citenamefont {Casas}\ and\ \citenamefont
  {Ibarra}(2001)}]{Casas:2001sr}%
  \BibitemOpen
  \bibfield  {author} {\bibinfo {author} {\bibfnamefont {J.}~\bibnamefont
  {Casas}}\ and\ \bibinfo {author} {\bibfnamefont {A.}~\bibnamefont {Ibarra}},\
  }\href {\doibase 10.1016/S0550-3213(01)00475-8} {\bibfield  {journal}
  {\bibinfo  {journal} {Nucl.Phys.}\ }\textbf {\bibinfo {volume} {B618}},\
  \bibinfo {pages} {171} (\bibinfo {year} {2001})},\ \Eprint
  {http://arxiv.org/abs/hep-ph/0103065} {arXiv:hep-ph/0103065 [hep-ph]}
  \BibitemShut {NoStop}%
%%CITATION = HEP-PH/0103065;%%
\bibitem [{\citenamefont {Maki}\ \emph {et~al.}(1962)\citenamefont {Maki},
  \citenamefont {Nakagawa},\ and\ \citenamefont {Sakata}}]{Maki:1962mu}%
  \BibitemOpen
  \bibfield  {author} {\bibinfo {author} {\bibfnamefont {Z.}~\bibnamefont
  {Maki}}, \bibinfo {author} {\bibfnamefont {M.}~\bibnamefont {Nakagawa}}, \
  and\ \bibinfo {author} {\bibfnamefont {S.}~\bibnamefont {Sakata}},\ }\href
  {\doibase 10.1143/PTP.28.870} {\bibfield  {journal} {\bibinfo  {journal}
  {Prog.Theor.Phys.}\ }\textbf {\bibinfo {volume} {28}},\ \bibinfo {pages}
  {870} (\bibinfo {year} {1962})}\BibitemShut {NoStop}%
%%CITATION = PTPKA,28,870;%%
\bibitem [{\citenamefont {'t~Hooft}(1980)}]{'tHooft:1979bh}%
  \BibitemOpen
  \bibfield  {author} {\bibinfo {author} {\bibfnamefont {G.}~\bibnamefont
  {'t~Hooft}},\ }\href@noop {} {\bibfield  {journal} {\bibinfo  {journal} {NATO
  Adv.Study Inst.Ser.B Phys.}\ }\textbf {\bibinfo {volume} {59}},\ \bibinfo
  {pages} {135} (\bibinfo {year} {1980})}\BibitemShut {NoStop}%
%%CITATION = NASBD,59,135;%%
\bibitem [{\citenamefont {Barbieri}\ and\ \citenamefont
  {Giudice}(1988)}]{Barbieri:1987fn}%
  \BibitemOpen
  \bibfield  {author} {\bibinfo {author} {\bibfnamefont {R.}~\bibnamefont
  {Barbieri}}\ and\ \bibinfo {author} {\bibfnamefont {G.}~\bibnamefont
  {Giudice}},\ }\href {\doibase 10.1016/0550-3213(88)90171-X} {\bibfield
  {journal} {\bibinfo  {journal} {Nucl.Phys.}\ }\textbf {\bibinfo {volume}
  {B306}},\ \bibinfo {pages} {63} (\bibinfo {year} {1988})}\BibitemShut
  {NoStop}%
%%CITATION = NUPHA,B306,63;%%
\bibitem [{\citenamefont {Asaka}\ \emph {et~al.}(2011)\citenamefont {Asaka},
  \citenamefont {Eijima},\ and\ \citenamefont {Ishida}}]{Asaka:2011pb}%
  \BibitemOpen
  \bibfield  {author} {\bibinfo {author} {\bibfnamefont {T.}~\bibnamefont
  {Asaka}}, \bibinfo {author} {\bibfnamefont {S.}~\bibnamefont {Eijima}}, \
  and\ \bibinfo {author} {\bibfnamefont {H.}~\bibnamefont {Ishida}},\ }\href
  {\doibase 10.1007/JHEP04(2011)011} {\bibfield  {journal} {\bibinfo  {journal}
  {JHEP}\ }\textbf {\bibinfo {volume} {1104}},\ \bibinfo {pages} {011}
  (\bibinfo {year} {2011})},\ \Eprint {http://arxiv.org/abs/1101.1382}
  {arXiv:1101.1382 [hep-ph]} \BibitemShut {NoStop}%
%%CITATION = ARXIV:1101.1382;%%
\bibitem [{\citenamefont {Besak}\ and\ \citenamefont
  {Bodeker}(2012)}]{Besak:2012qm}%
  \BibitemOpen
  \bibfield  {author} {\bibinfo {author} {\bibfnamefont {D.}~\bibnamefont
  {Besak}}\ and\ \bibinfo {author} {\bibfnamefont {D.}~\bibnamefont
  {Bodeker}},\ }\href {\doibase 10.1088/1475-7516/2012/03/029} {\bibfield
  {journal} {\bibinfo  {journal} {JCAP}\ }\textbf {\bibinfo {volume} {1203}},\
  \bibinfo {pages} {029} (\bibinfo {year} {2012})},\ \Eprint
  {http://arxiv.org/abs/1202.1288} {arXiv:1202.1288 [hep-ph]} \BibitemShut
  {NoStop}%
%%CITATION = ARXIV:1202.1288;%%
\bibitem [{\citenamefont {Harvey}\ and\ \citenamefont
  {Turner}(1990)}]{Harvey:1990qw}%
  \BibitemOpen
  \bibfield  {author} {\bibinfo {author} {\bibfnamefont {J.~A.}\ \bibnamefont
  {Harvey}}\ and\ \bibinfo {author} {\bibfnamefont {M.~S.}\ \bibnamefont
  {Turner}},\ }\href {\doibase 10.1103/PhysRevD.42.3344} {\bibfield  {journal}
  {\bibinfo  {journal} {Phys.Rev.}\ }\textbf {\bibinfo {volume} {D42}},\
  \bibinfo {pages} {3344} (\bibinfo {year} {1990})}\BibitemShut {NoStop}%
%%CITATION = PHRVA,D42,3344;%%
\bibitem [{\citenamefont {Cui}\ \emph {et~al.}(2012)\citenamefont {Cui},
  \citenamefont {Randall},\ and\ \citenamefont {Shuve}}]{Cui:2011ab}%
  \BibitemOpen
  \bibfield  {author} {\bibinfo {author} {\bibfnamefont {Y.}~\bibnamefont
  {Cui}}, \bibinfo {author} {\bibfnamefont {L.}~\bibnamefont {Randall}}, \ and\
  \bibinfo {author} {\bibfnamefont {B.}~\bibnamefont {Shuve}},\ }\href
  {\doibase 10.1007/JHEP04(2012)075} {\bibfield  {journal} {\bibinfo  {journal}
  {JHEP}\ }\textbf {\bibinfo {volume} {1204}},\ \bibinfo {pages} {075}
  (\bibinfo {year} {2012})},\ \Eprint {http://arxiv.org/abs/1112.2704}
  {arXiv:1112.2704 [hep-ph]} \BibitemShut {NoStop}%
%%CITATION = ARXIV:1112.2704;%%
\bibitem [{\citenamefont {Ade}\ \emph {et~al.}(2013)\citenamefont {Ade} \emph
  {et~al.}}]{Ade:2013zuv}%
  \BibitemOpen
  \bibfield  {author} {\bibinfo {author} {\bibfnamefont {P.}~\bibnamefont
  {Ade}} \emph {et~al.} (\bibinfo {collaboration} {Planck Collaboration}),\
  }\href@noop {} {\  (\bibinfo {year} {2013})},\ \Eprint
  {http://arxiv.org/abs/1303.5076} {arXiv:1303.5076 [astro-ph.CO]} \BibitemShut
  {NoStop}%
%%CITATION = ARXIV:1303.5076;%%
\bibitem [{\citenamefont {Glashow}\ and\ \citenamefont
  {Weinberg}(1977)}]{Glashow:1976nt}%
  \BibitemOpen
  \bibfield  {author} {\bibinfo {author} {\bibfnamefont {S.~L.}\ \bibnamefont
  {Glashow}}\ and\ \bibinfo {author} {\bibfnamefont {S.}~\bibnamefont
  {Weinberg}},\ }\href {\doibase 10.1103/PhysRevD.15.1958} {\bibfield
  {journal} {\bibinfo  {journal} {Phys.Rev.}\ }\textbf {\bibinfo {volume}
  {D15}},\ \bibinfo {pages} {1958} (\bibinfo {year} {1977})}\BibitemShut
  {NoStop}%
%%CITATION = PHRVA,D15,1958;%%
\bibitem [{CMS(2013{\natexlab{a}})}]{CMS-PAS-HIG-13-004}%
  \BibitemOpen
  \href@noop {} {\emph {\bibinfo {title} {{Search for the Standard-Model Higgs
  boson decaying to tau pairs in proton-proton collisions at sqrt(s) = 7 and 8
  TeV}}}},\ \bibinfo {type} {Tech. Rep.}\ \bibinfo {number}
  {CMS-PAS-HIG-13-004}\ (\bibinfo  {institution} {CERN},\ \bibinfo {address}
  {Geneva},\ \bibinfo {year} {2013})\BibitemShut {NoStop}%
\bibitem [{\citenamefont {Peskin}(2012)}]{Peskin:2012we}%
  \BibitemOpen
  \bibfield  {author} {\bibinfo {author} {\bibfnamefont {M.~E.}\ \bibnamefont
  {Peskin}},\ }\href@noop {} {\  (\bibinfo {year} {2012})},\ \Eprint
  {http://arxiv.org/abs/1207.2516} {arXiv:1207.2516 [hep-ph]} \BibitemShut
  {NoStop}%
%%CITATION = ARXIV:1207.2516;%%
\bibitem [{\citenamefont {Eriksson}\ \emph {et~al.}(2010)\citenamefont
  {Eriksson}, \citenamefont {Rathsman},\ and\ \citenamefont
  {Stal}}]{Eriksson:2009ws}%
  \BibitemOpen
  \bibfield  {author} {\bibinfo {author} {\bibfnamefont {D.}~\bibnamefont
  {Eriksson}}, \bibinfo {author} {\bibfnamefont {J.}~\bibnamefont {Rathsman}},
  \ and\ \bibinfo {author} {\bibfnamefont {O.}~\bibnamefont {Stal}},\ }\href
  {\doibase 10.1016/j.cpc.2009.09.011} {\bibfield  {journal} {\bibinfo
  {journal} {Comput.Phys.Commun.}\ }\textbf {\bibinfo {volume} {181}},\
  \bibinfo {pages} {189} (\bibinfo {year} {2010})},\ \Eprint
  {http://arxiv.org/abs/0902.0851} {arXiv:0902.0851 [hep-ph]} \BibitemShut
  {NoStop}%
%%CITATION = ARXIV:0902.0851;%%
\bibitem [{\citenamefont {Liu}\ \emph {et~al.}(2013)\citenamefont {Liu},
  \citenamefont {Shuve}, \citenamefont {Weiner},\ and\ \citenamefont
  {Yavin}}]{Liu:2013gba}%
  \BibitemOpen
  \bibfield  {author} {\bibinfo {author} {\bibfnamefont {J.}~\bibnamefont
  {Liu}}, \bibinfo {author} {\bibfnamefont {B.}~\bibnamefont {Shuve}}, \bibinfo
  {author} {\bibfnamefont {N.}~\bibnamefont {Weiner}}, \ and\ \bibinfo {author}
  {\bibfnamefont {I.}~\bibnamefont {Yavin}},\ }\href {\doibase
  10.1007/JHEP07(2013)144} {\bibfield  {journal} {\bibinfo  {journal} {JHEP}\
  }\textbf {\bibinfo {volume} {1307}},\ \bibinfo {pages} {144} (\bibinfo {year}
  {2013})},\ \Eprint {http://arxiv.org/abs/1303.4404} {arXiv:1303.4404
  [hep-ph]} \BibitemShut {NoStop}%
%%CITATION = ARXIV:1303.4404;%%
\bibitem [{CMS(2013{\natexlab{b}})}]{CMS-PAS-SUS-13-006}%
  \BibitemOpen
  \href@noop {} {\emph {\bibinfo {title} {{Search for electroweak production of
  charginos, neutralinos, and sleptons using leptonic final states in pp
  collisions at 8 TeV}}}},\ \bibinfo {type} {Tech. Rep.}\ \bibinfo {number}
  {CMS-PAS-SUS-13-006}\ (\bibinfo  {institution} {CERN},\ \bibinfo {address}
  {Geneva},\ \bibinfo {year} {2013})\BibitemShut {NoStop}%
\bibitem [{\citenamefont {Kanemura}\ \emph {et~al.}(2012)\citenamefont
  {Kanemura}, \citenamefont {Tsumura},\ and\ \citenamefont
  {Yokoya}}]{Kanemura:2011kx}%
  \BibitemOpen
  \bibfield  {author} {\bibinfo {author} {\bibfnamefont {S.}~\bibnamefont
  {Kanemura}}, \bibinfo {author} {\bibfnamefont {K.}~\bibnamefont {Tsumura}}, \
  and\ \bibinfo {author} {\bibfnamefont {H.}~\bibnamefont {Yokoya}},\ }\href
  {\doibase 10.1103/PhysRevD.85.095001} {\bibfield  {journal} {\bibinfo
  {journal} {Phys.Rev.}\ }\textbf {\bibinfo {volume} {D85}},\ \bibinfo {pages}
  {095001} (\bibinfo {year} {2012})},\ \Eprint {http://arxiv.org/abs/1111.6089}
  {arXiv:1111.6089 [hep-ph]} \BibitemShut {NoStop}%
%%CITATION = ARXIV:1111.6089;%%
\bibitem [{ATL(2013)}]{ATLAS-CONF-2013-049}%
  \BibitemOpen
  \href@noop {} {\emph {\bibinfo {title} {{Search for direct-slepton and
  direct-chargino production in final states with two opposite-sign leptons,
  missing transverse momentum and no jets in 20/fb of pp collisions at sqrt(s)
  = 8 TeV with the ATLAS detector}}}},\ \bibinfo {type} {Tech. Rep.}\ \bibinfo
  {number} {ATLAS-CONF-2013-049}\ (\bibinfo  {institution} {CERN},\ \bibinfo
  {address} {Geneva},\ \bibinfo {year} {2013})\BibitemShut {NoStop}%
\bibitem [{\citenamefont {Abbiendi}\ \emph {et~al.}(2004)\citenamefont
  {Abbiendi} \emph {et~al.}}]{Abbiendi:2003ji}%
  \BibitemOpen
  \bibfield  {author} {\bibinfo {author} {\bibfnamefont {G.}~\bibnamefont
  {Abbiendi}} \emph {et~al.} (\bibinfo {collaboration} {OPAL Collaboration}),\
  }\href {\doibase 10.1140/epjc/s2003-01466-y} {\bibfield  {journal} {\bibinfo
  {journal} {Eur.Phys.J.}\ }\textbf {\bibinfo {volume} {C32}},\ \bibinfo
  {pages} {453} (\bibinfo {year} {2004})},\ \Eprint
  {http://arxiv.org/abs/hep-ex/0309014} {arXiv:hep-ex/0309014 [hep-ex]}
  \BibitemShut {NoStop}%
%%CITATION = HEP-EX/0309014;%%
\end{thebibliography}%
\end{document}